\documentclass[aps,pre,floats,floatfix,twocolumn,longbibliography]{revtex4-1}

\usepackage{graphicx}
\usepackage{epstopdf}
\graphicspath{{./Figs/}{./figs/}}

\usepackage{latexsym}
\usepackage{amsmath}
\usepackage{amssymb}
\usepackage{bm}
\usepackage{wasysym}

\usepackage{color}
\usepackage[colorlinks,allcolors=blue,bookmarks=true]{hyperref}
\usepackage{flafter}
\usepackage[percent]{overpic}

\newcommand{\const}{\mbox{const}}

\newcommand{\im}{\mbox{Im}}
\newcommand{\re}{\mbox{Re}}

\newcommand{\amatrix}[1]{\begin{matrix} #1 \end{matrix}}

\newcommand{\bra}[1]{\left\langle #1 \right|}
\newcommand{\ket}[1]{\left| #1 \right\rangle}
\newcommand{\braket}[1]{\left\langle #1 \right\rangle}

\newcommand{\BraKet}[3]{\left\langle #1 \middle| #2 \middle| #3 \right\rangle}

\newcommand{\be}[1]{\begin{eqnarray}{\label{e#1}}} 
\newcommand{\beq}{\begin{eqnarray}}
\newcommand{\eeq}{\end{eqnarray}} 

\newcommand{\hide}[1]{}
\newcommand{\Eq}[1]{\textcolor{blue}{{Eq.}\!\!~(\ref{#1})}}

\newcommand{\Fig}[1]{\textcolor{blue}{Fig.}\!\!~\ref{#1}} 

\definecolor{myred}{rgb}  {0.5,0.0,0.0}

\newcommand{\hrefl}[2]{\href{#2}{(#1)}}
\newcommand{\sect}[1]{{\bf #1.--}}

\begin{document}

\title{Negative mobility, sliding and delocalization for stochastic networks}

\author{Dima Boriskovsky, Doron Cohen}

\affiliation{\mbox{Department of Physics, Ben-Gurion University of the Negev, Beer-Sheva 84105, Israel}}

\begin{abstract}
We consider prototype configurations for quasi-one-dimensional stochastic networks that exhibit negative mobility, meaning that current decreases or even reversed as the bias is increased. We then explore the implications of disorder. In particular we ask whether lower and upper bias thresholds restrict the possibility to witness non-zero current (sliding and anti-sliding transitions respectively), and whether a delocalization effect manifest itself (crossover from over-damped to under-damped relaxation). In the latter context detailed analysis of the relaxation spectrum as a function of the bias is provided for both on-chain and off-chain disorder.     
\end{abstract}

\maketitle

\section{Introduction}

Negative mobility, where a system respond to the bias in an opposite way to the naive expectation, has been studied experimentally long ago for semiconductors, diodes, and superlattices that feature resonance tunneling \cite{esaki1,esaki2,sibille,keay} or carriers with negative effective mass \cite{dousmanis,kromer,lei,cannon}.  
But it has been realized that such effect can be expected also for a purely stochastic {\em hopping} conductance in the presence of high electric field~\cite{bottger}.    
The physics involved is that of biased diffusion in random structures such as percolating systems above criticality. The essential physics is captured by simpler quasi-one-dimensional configurations, notably by {\em comb or tree structures} that feature random distribution of dangling branches or waiting times \cite{havlin,balakrishnan,zia,benarous} (and see further references therein). Negative mobility can be described as ``getting less from pushing more", and there is a possibility of observing an upper critical bias beyond which the drift velocity vanishes.
More complicated configurations have been considered as well, for example: the flow of particles through a narrow tube with hooks that provide trapping mechanism~\cite{baerts}; inertial tracers in steady laminar flows~\cite{sarracino}; kinetically constrained systems~\cite{cividini,turci}; possibly involving several species of carriers~\cite{reichhardt} or mixture of gases~\cite{vrhovac}. 
Most of the cited examples above refer to negative differential mobility (NDM), but some also to absolute negative mobility (ANM), notably Brownian motors \cite{reimann,hanggi2002,hanggi}.  The latter are required to be non-equilibrium, or active systems in some sense \cite{cleuren,ndm}. We further illuminate the latter observation below.

\begin{figure}[b]
\centering
\includegraphics[width=0.84\hsize]{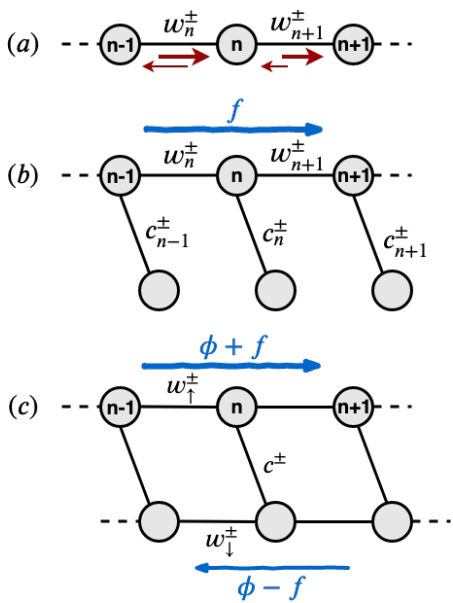}   
\caption{
{\bf Schematic drawing of the model system.} 
(a)~The standard Sinai model, namely, one dimensional chain where the transitions-rates have independent random values.  
(b)~One dimensional chain with dangling bonds that serves as a minimal model for a stochastic percolating network. 
(c)~Quasi one dimensional network that serves as a minimal model for an active gas of Janus-type particles. 
In panel~a the rates are indicated by arrows. Such arrows are omitted in panels b-c where we prefer to indicate the preferred direction of transitions due to the external bias and the self-propulsion. In the absence of disorder the model parameters in panel~a are the transition rate~$w_o$ for unbiased transitions, and the bias field~$f$ that encourages motion to the right. Sinai disorder means that~$f$ acquires an uncorrelated random component. In panel~b we have additional dangling bonds. If those transitions rates are random, we call it ``off chain disorder". In panel~c we also have propulsion~$\phi$ that encourages the motion to be in a direction that agrees with the orientation of the particle. 
See text for further details, and Eqs.(2-9) for precise definition of the model parameters.} 
\label{f1}
\end{figure}

\sect{Objective}
In the present work we consider {\em minimal} stochastic models where either NDM or ANM can be expected. Those are illustrated in \Fig{f1}. In both cases the configuration reflects the existence of more than one dimension, as opposed to a simple one dimensional chain with near-neighbour transitions. The NDM configuration of \Fig{f1}b is a minimal comb structure of the type that has been studied in the past \cite{havlin,balakrishnan,zia,cleuren,benarous}. The ANM configuration of \Fig{f1}c is a simplified  version of \cite{cleuren} that has been studied in \cite{ndm}. The ANM configuration is further characterized by a non-trivial topology. Namely, it is an active stochastic network that features loops, around which the circulation of the stochastic field is non-zero. It can represent the dynamics of a Janus particle \cite{janus1,janus2,janus3} in one dimension \cite{ndm,kafri}, where the extra degree of freedom is its orientation. 
Our main interest concerns the implication of disorder on the stochastic relaxation. We provide below some background to the relevant literature, followed by an outline that explains our motivation to further study the prototype configuration of \Fig{f1}b, which complements our previous studies \cite{psl,ndm} of relaxation for the configurations \Fig{f1}a and \Fig{f1}c respectively.

\sect{Sinai model}
The study of stochastic motion on a one dimensional random lattice with near neighbour transitions has been introduced by Sinai \cite{sinai}, aka {\em random walk in random environment}. The model is illustrated in \Fig{f1}a. Unlike Einstein's Brownian motion, here the rates of transition between two adjacent sites do not have to be the same in both directions. This can be regarded as arising from a {\em stochastic field}, due to a potential difference that biases the transitions. If the stochastic field is uncorrelated on different bonds, it follows that the potential is characterized by an activation barrier whose hight scales as $\sqrt{L}$, where $L$ is the length of the sample. It follows that the steady state current is ${I \sim \exp[-\const \sqrt{L}]}$, and not ${I \propto 1/L}$. In time domains it implies sub-diffusion $R \sim \ln^2(t)$, where $R$ is the distance that is covered by the particle during time~$t$.
This differs completely from the usual random walk result $R \sim t^{1/2}$.

\sect{Sliding transition}
Adding bias in the context of Sinai model means that the rates in one direction (say to the right) are, on the average, larger than the rates of transition in the opposite direction. It turns out that the system exhibits a non-zero drift velocity $v$ provided the bias exceeds a finite critical value. This is known as the {\em sliding transition} \cite{derrida}. For sub-critical bias the drift is $R \sim t^{\mu}$, where ${\mu \in [0,1]}$ depends on the bias. For a comprehensive review see \cite{bouchaud1,bouchaud2}. Here we focus on the possibility that above some second critical bias the drift velocity becomes zero again, relating to the minimal configuration of \Fig{f1}b.

\sect{Delocalization transition}
Somewhat related to the sliding transition, is the delocalization transition of the relaxation modes. Here one considers a ring geometry: a chain segment with periodic boundary conditions. The delocalization transition has been discussed originally for non-Hermitian Hamiltonians \cite{nelson1,nelson2,nelson3}, and only later for stochastic chains \cite{rss}. In the next paragraph we explain the term delocalization in the latter context. 

In the presence of bias (aka affinity), due to the disorder, some (or all) eigenvalues become real, which is regarded as an indication for the localization of the associated eigenmodes. As the bias is increased, some of those real eigenvalues become complex, which is termed delocalization transition. 
The delocalization of the eigenmodes whose eigenvalues reside in the vicinity of $\lambda{=}0$ implies that the relaxation become under-damped (with oscillations).
As explained in \cite{rss}, the threshold for that is lower than the threshold for the sliding transition (it corresponds to $\mu{=}1/2$ and not to $\mu{=}1$). 

The delocalization of the relaxation modes for the active network of \Fig{f1}c has been already studied in \cite{ndm}. Here we focus on the configuration of \Fig{f1}b, and distinguish between on-chain disorder and off-chain disorder. This part of the study is motivated by the following question: we know that in one dimension we always have localization; does it mean that in a closed ring we always have a delocalization transition?

\sect{Outline}
We highlight the theme of negative mobility (NDM/ANM) for {\em minimal} quasi one dimensional networks, with emphasis on the distinction between on-chain and off-chain disorder. The preliminary sections are mainly pedagogical, and clarify the dependence of the steady state current on the bias. This prepares the grounds to the main sections, that expand on the relaxation spectrum and the delocalization transition. The detailed outline is as follows:      
{\bf (1)}~In Section~II we elaborate on the NDM/ANM configurations of \Fig{f1}bc. The model parameters are defined in Eqs.(2-9).
{\bf (2)}~In Sections~III and~IV we derive expressions for the dependence of the current on the bias for a non-disordered chain, and illustrate the NDM/ANM effect. 
{\bf (3)}~In Section~IV we consider the effect of disorder on current-vs-bias for the network of \Fig{f1}b. This section highlight the possiblity to observe an anti-sliding transition, namely, suppression of the current for bias that exceeds a threshold. 
{\bf (4)}~In Section~V we find analytically the relaxation spectrum for a chain that is closed into a non-disordered ring configuration. From that we can re-derive the analytical result for the current, and additionally we get an analytical result for the diffusion coefficient.     
{\bf (5)}~In Section VI we consider the effect of disorder. We explore the possibility to observe an over-damped relaxation and a delocalization transition. We distinguish between on-chain and off-chain disorder. 
{\bf (6)}~In Section VII we further analyze analytically the localization of the relaxation modes, via a reduction to the Anderson-Debye model. 
A few appendices provide extra technical details. The main results are summarized in Section~VIII.

\section{The model} 

The dynamics of the stochastic particle is described by a rate equation  
\beq
\frac{d}{dt}\mathbf{p} \ \ = \ \ \bm{\mathcal{W}} \mathbf{p}
\eeq
where $\mathbf{p}$ is a vector of probabilities and $\bm{\mathcal{W}}$ is a matrix of transition rates. The off-diagonal element $w_{n,m}$ of the matrix is the transition-rate from node~$m$ to node~$n$. The diagonal elements ${-\gamma_n}$ are determined such that the sum over each column is zero. Accordingly $\gamma_n$ is the total decay-rate from node $n$.

We consider a {\em quasi} one-dimensional chains, as in \Fig{f1}. Therefore we have to labels the nodes of the network by a composite index. Namely, the nodes are labelled by a site index ${n=\text{integer}}$ and by an additional index ${s=\uparrow,\downarrow}$. 
For a non-disordered chain, by {\em convention}, all the forward rates to right are $w_o$ along the chain, 
and all the outward rates to the dangling sites are $c_o$. We assume bias $f$, 
and {\em define} the backwards transition rates as $w_o e^{-f}$ and $c_o e^{-\alpha f}$ respectively, 
where ${\alpha>0}$ is a proportionality constant that quantifies the relative sensitivity of the dangling bonds to the bias. We emphasize that without loss of generality 
the forward rates, by this {\em convention}, are not affected by the bias. 
This helps to maintain a numerically meaningful ${f\rightarrow\infty}$ limit.    

\Fig{f1}b describes a one-dimensional lattice with dangling bonds.
The non zero transition rates are: 
\beq  
w_{n\uparrow ,n{-}1 \uparrow} \ &=& \ w_n^{+} \ \ := \ \ w_o  \\
w_{n{-}1\uparrow ,n \uparrow} \ &=& \ w_n^{-}  \ \ := \ \ w_o e^{-f} \\
\label{eCp} w_{n\downarrow ,n \uparrow} \ \ \ \ &=& \ c_n^{+}\ \ \ := \ \  c_o  \\
\label{eCm} w_{n\uparrow ,n \downarrow} \ \ \ \ &=& \ c_n^{-} \ \ \ := \ \ c_o e^{-\alpha f} 
\eeq
The last expression in each row refers to a non-disordered sample.
Note that in our convention bond $n$ connects node $n$ to node ${n{-}1}$.  
The $s$ index can represent a transverse space coordinate, 
or the possibility of the particle to switch into a non-conducting state.

\begin{figure}[b!] 
\centering
\includegraphics[width=\hsize]{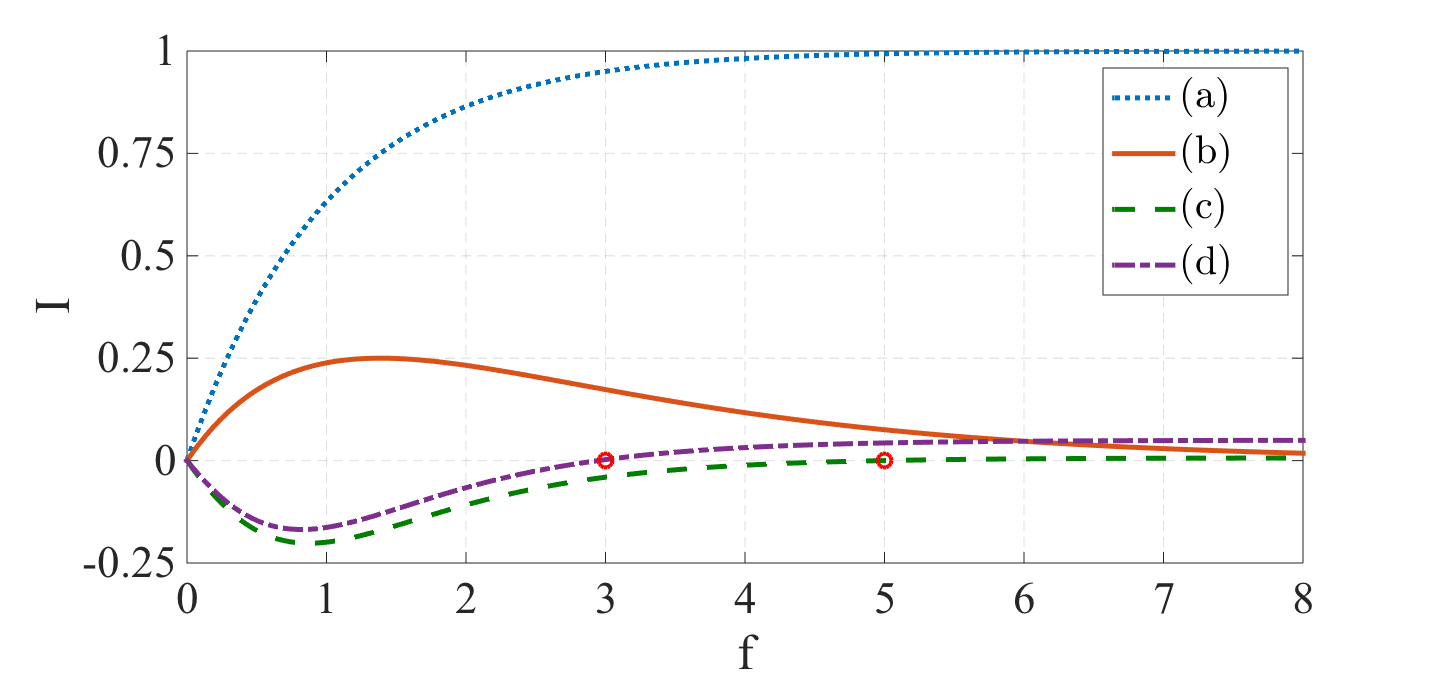} 
\includegraphics[width=\hsize]{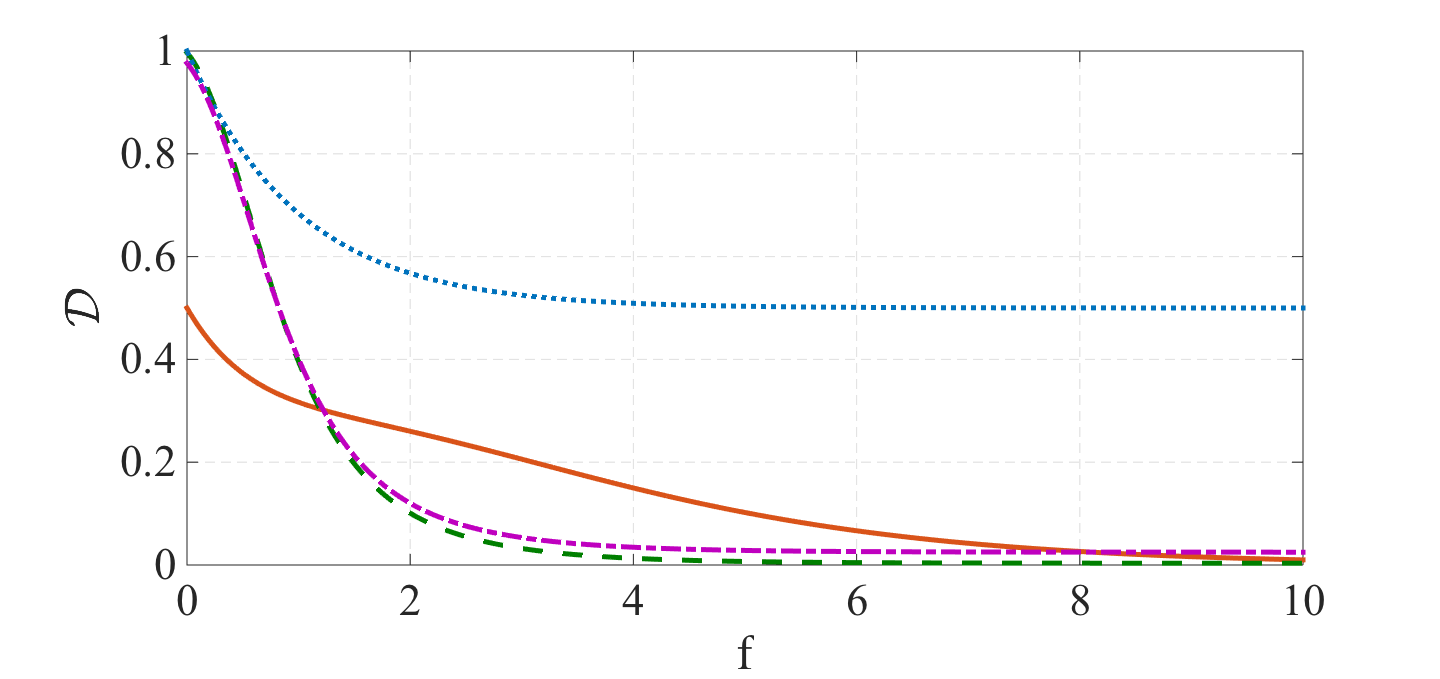} 
\caption{{\bf Current and diffusion versus bias.} 
The {\em upper} panel displays the current in units of $w_o/L$, 
plotted versus the bias~$f$. The curves are for:  
(a)~Simple chain with no dangling bonds; 
(b)~Chain with dangling bonds ($\alpha=1/2$).  
(c)~Active chain with $\alpha=2$ and $\phi=5$. 
(d)~Active chain with $\alpha=2$ and $\phi=3$.
For the active chain ${c_o=w_o}$,  
and the red circles indicate the current reversal.
The {\em lower} panel displays, respectively,  
the diffusion coefficient $\mathcal{D}$ in units of $w_o$.}  
\label{f2}
\end{figure}

\Fig{f1}c describes a Janus (active) particle that can have forward or backward orientation: in the $\uparrow$ orientation it executes a stochastic self-propelled motion that is biased to the right, while in the $\downarrow$ orientation it executes a stochastic self-propelled motion that is biased to the left. Due to self-propulsion there is some ratio ${\exp(\phi)>1}$ between the forward and the backward motion. By our {\em convention} only the backward rates (relative to the propulsion) are affected, hence the transition rates to the left in the upper edges become $w_o e^{-\phi-f}$, and the transition rates to the right in the lower edges become $w_o e^{-\phi}$. Summarizing, for a non-disordered motion of a Janus particle   
\beq  
w_{n\uparrow ,n{-}1 \uparrow} \ &=& \ w_{\uparrow}^{+} \ \ = \ \ w _o  \\
w_{n{-}1\uparrow ,n \uparrow} \ &=& \ w_{\uparrow}^{-}  \ \ = \ \ w_o e^{-f-\phi} \\
w_{n\downarrow ,n{-}1 \downarrow} \ &=& \ w_{\downarrow}^{+}  \ \ = \ \ w _o e^{-\phi} \\
w_{n{-}1\downarrow ,n \downarrow} \ &=& \ w_{\downarrow}^{-} \ \ = \ \ w _o  e^{-f} 
\eeq    
while the $c^{\pm}$ are given by \Eq{eCp} and \Eq{eCm}.

Different types of disorder can be introduced as discussed thoroughly for a simple chain \cite{psl}, 
and for the configuration of \Fig{f1}c \cite{ndm}.
In the present work, referring to \Fig{f1}b, the interesting distinction 
is between on-chain disorder and off-chain disorder.
On-chain disorder means that the bias field $f$ has uncorrelated values $f_n$ on the different bonds of the chain, as in the standard Sinai model. Specifically we assume box distribution with average value~$f$ that reflects the presence of an externally applied bias, plus a random component due to the embedding environment. Accordingly we write   
\beq \label{eOCD}
f_n \ \ = \ \ f + \text{random}[-\sigma,\sigma]    
\eeq
Off-chain disorder is similarly defined. Optionally we can regard it as arising from random $\alpha$. Namely, in the presence of an external bias~$f$, the stochastic field on the $n$-th dangling bond is $\alpha_n f$ with random values for $\alpha_n$. Otherwise, we can specify separately $\sigma_{\text{on}}$ for the random field on the chain bonds, and $\sigma_{\text{off}}$ for the random field on the dangling bonds. Which convention is used is a matter of context, per the assumed physical-setup.

\section{The steady state current for a non-disordered chain}

The non-equilibrium steady state (NESS) is determined by the equation $\bm{\mathcal{W}}\mathbf{p}=0$, which is formally a continuity equation. For the prototype percolating network of \Fig{f1}b, the drift velocity along the chain sites is 
\beq
v_{\uparrow} \ \ = \ \ (1-e^{-f}) w_o
\eeq
Without dangling bonds the occupation probability at each site of the chain is ${p_{\text{chain}}=1/L}$, where $L$ is the length of the chain, hence the current is $I=(1/L)v_{\uparrow}$. With added dangling bonds the NESS equation 
implies ${p_{\downarrow}/p_{\uparrow} = e^{\alpha f}}$, hence 
\beq
p_{\uparrow} \ \ = \ \ \frac{1}{L} \left( \frac{1}{1 + e^{\alpha f}} \right)
\eeq
and accordingly 
\beq \label{e4}
I \ \ = \ \ p_{\uparrow} v_{\uparrow} \ \ =  \ \ \frac{w_o}{L} \left( \frac{1 - e^{-f}}{1 + e^{\alpha f}} \right)
\eeq
For the network of \Fig{f1}c the same NESS occupation prevails, but the current becomes  
\beq 
I \ \ &=& \ p_{\uparrow} v_{\uparrow} -  p_{\downarrow} v_{\downarrow}   \\
&=& \ \ \frac{w_o}{L} \left( \frac{[1 - e^{-\phi-f}] + e^{\alpha f}[e^{-\phi}-e^{-f}] }{1 + e^{\alpha f}} \right)
\eeq
For small $f$ we get a linear relation ${I \approx Gf}$, with 
\beq
G \ \ = \ \ \frac{w_o}{2L} \left[ (1{+}\alpha)e^{-\phi} + (1{-}\alpha) \right]
\eeq
We conclude that ANM will show up if ${\alpha>1}$ provided the propulsion 
is strong enough, namely, 
\beq
\phi \ \ >  \ \ \ln\left(\frac{\alpha{+}1}{\alpha{-}1}\right)
\eeq
Several $I$ versus $f$ plots are displayed in \Fig{f2}a.
For completeness we also plot the diffusion coefficient $\mathcal{D}$ in \Fig{f2}b.
The way to calculate $\mathcal{D}$ will be explained in Section~V.

\begin{figure}[b!] 
\centering
\includegraphics[width=\hsize]{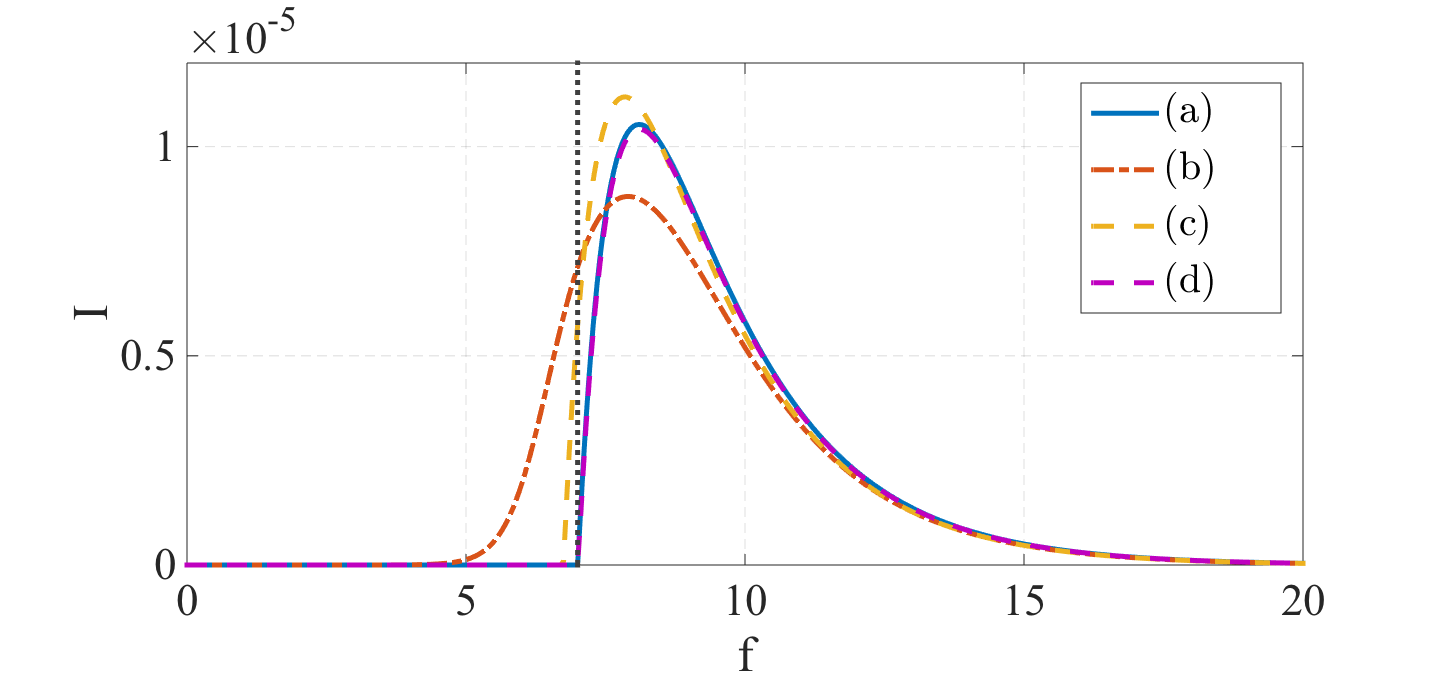} 
\caption{
{\bf Sliding transition for a chain with dangling bonds.} 
The current in units of $w_o/L$ is plotted versus the bias $f$ with $\sigma=10$ and ${\alpha=1/2}$.
(a)~Analytical curve plotted by \Eq{e15}.
(b)~Numerical results with a realization of $L=35$.
(c)~Analytical results evaluated by \Eq{e14} with the same realization as b.
(d)~Analytical results evaluated by \Eq{e14} with $L=25,000$.
The dotted line is $f_{s} \approx 7$. }
\label{f3}
\includegraphics[width=\hsize]{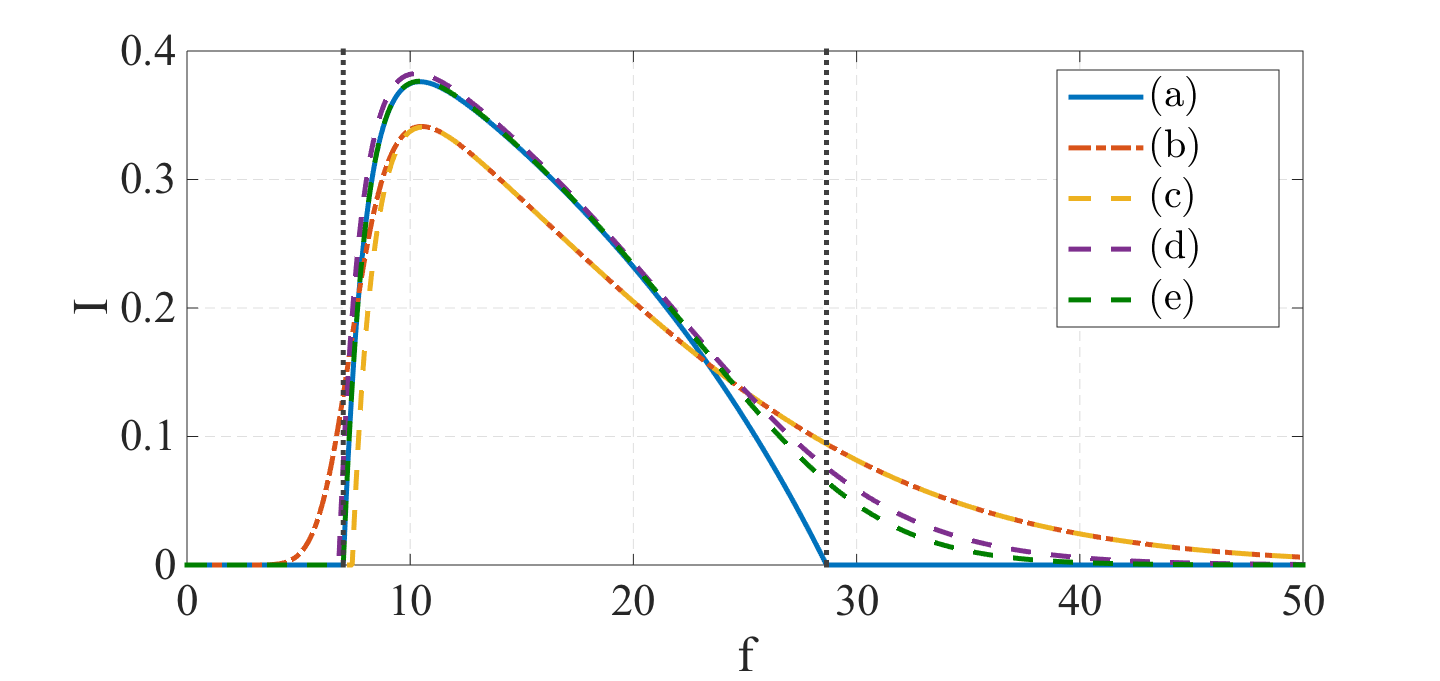} 
\caption{
{\bf Anti sliding transitions for a chain with stretched distribution of dangling bonds.}
The current in units of $w_o/L$ is plotted versus the bias $f$ with $\sigma=10$ and $\alpha=0.035$.
(a)~Analytical curve plotted by \Eq{e18}.
(b)~Numerical results with a realization of $L=35$.
(c)~Analytical results evaluated by \Eq{e14} with the same realization as b. 
(d)~Analytical results evaluated by \Eq{e14} with $L=500$. 
(e)~Analytical results evaluated by \Eq{e14} with $L=25,000$. 
The dotted lines are $f_s \approx 7$ (left) and $f_c \approx 28.65$ (right). }
\label{f4}
\end{figure}

\section{The steady state current for a disordered chain} 

We consider again the network of \Fig{f1}b, but now with Sinai disorder. This means on-chain disorder that is given by \Eq{eOCD}, and an additional off-chain disorder (due to random $\alpha$) that will be specified later on. Our objective is to determine the steady state current~$I$ in the presence of disorder, and to see whether it diminishes if the bias is below or above some thresholds.   
Recall that the NESS is determined by the equation $\bm{\mathcal{W}}\mathbf{p}=0$, 
which is formally a continuity equation. Along the $n$-th dangling bond it is implied that
\beq 
\frac{p_{n,\downarrow}}{p_{n,\uparrow}} \ \  = \ \  e^{\alpha_n f}
\eeq
because the NESS current there has to be zero.  
Along the $n$-th bond of the chain we require 
\beq \label{eNESS} 
w_{n}^{+} p_{n-1,\uparrow} - w_{n}^{-} p_{n,\uparrow}  \ = \ I
\eeq
If one drops the $\downarrow$ dangling sites, 
this equation with the normalization $\sum_n  p_{n} =1$ leads
to a solution $p_n$, that is formally identical to the solution 
that has been obtained by Derrida \cite{derrida} for a simple ring. 
Namely, the current is $I=(1/L)v$, where 
\beq \label{eq:024}
v &=& \left( 1 - \braket{\frac{w_n^{-}}{w_n^{+}}}\right) \braket{\frac{1}{w_n^{+}}}^{-1} \\
&=& \left(1 - \frac{1}{L} \sum_n e^{-f_n} \right) w_o
\eeq
This expression is valid if it gives a non-negative result. 
It should be realized that $\braket{e^{-f_n}}$ is larger than unity 
if the $f_n$ have zero or small enough average.
Consequently, for small bias the above expression becomes negative, 
indicating that $v=0$ in the $L\rightarrow\infty$ limit. 
The transition from zero drift velocity to finite drift velocity 
as the bias exceeds a threshold is known as the {\em sliding transition} \cite{derrida,bouchaud1,bouchaud2}.

If we place back the dangling sites the solution of \Eq{eNESS} will 
be the same up to a factor, namely  
\beq
p_{n,\uparrow} = p_{\uparrow} p_n
\eeq
where $p_{\uparrow}$ is determined by the normalization condition 
\beq
\sum_{n,s} p_{n,s}  \ = \ \sum_n (1+e^{\alpha_n f}) \, p_{n,\uparrow} \ = \ 1
\eeq
The off-chain disorder $\alpha_n$ is assumed to be independent of the on-chain disorder. 
We therefore can factorize the ensemble average, and deduce that 
\beq   
p_{\uparrow} \ \ &=& \ \ \frac{1}{L} \left( \frac{1}{1+\braket{e^{\alpha_n f}}} \right)
\eeq
Consequently for the current we get 
\beq \label{e14}
I \ \ = \ \ p_{\uparrow} v \ \ =  \ \ \frac{w_o}{L} \left( \frac{1-\braket{e^{-f_n}}}{1+\braket{e^{\alpha_n f}}} \right)
\eeq
This expression still features the Derrida sliding transition, but it can also provide 
an anti-sliding transition for large bias, if $\braket{e^{\alpha_n f}}$ becomes infinite.
As in the case of the standard Sinai model \cite{sinai,derrida,bouchaud1,bouchaud2},
also here both the sliding and the anti-sliding transitions become sharp only in the limit of a large chain (${L\rightarrow\infty}$).

Both the sliding transition and the anti-sliding transition are demonstrated in \Fig{f3} and \Fig{f4}. In \Fig{f3} we assume that the $f_n$ are box distributed within ${[f-\sigma,f+\sigma]}$, and that the $\alpha_n f$ are similarly distributed within ${[\alpha f-\sigma,\alpha f+\sigma]}$, we get
\beq \label{e15}
I \ \ = \ \frac{w_o}{L} \left( \frac{1-[\sinh(\sigma)/\sigma]e^{-f}}{1+ [\sinh(\sigma)/\sigma] e^{\alpha f}} \right)
\eeq
Here we have only a sliding transition because the denominator does not diverge for finite bias.
In \Fig{f4} we assume a stretched distribution for the dangling bonds, 
namely, we assume that the $\alpha_n f$ have exponential distribution 
with an average $\alpha f$. Accordingly 
\beq
\braket{e^{\alpha_n f}} \
&=& \ \frac{1}{\alpha f} \int_{0}^{\infty} e^{f'}e^{-f'/(\alpha f)} df' \\
&=& \  
\begin{cases}
\frac{1}{1{-}\alpha f} & \alpha f<1 \\
\infty & \alpha f>1
\end{cases}
\eeq 
Here the expectation values diverges for large $f$, 
and therefore for larger $f$ the current vanishes.
At the regime where the current is finite we get 
\beq \label{e18}
I \ \ = \ \frac{w_o}{L} \left[\frac{1 - \alpha f}{2-\alpha f}\right] \left(1- [\sinh(\sigma)/\sigma] e^{-f} \right)
\eeq
This expression gives non-zero result 
within the range ${f_s < f  < f_c }$, where 
\beq
f_s \ &=& \ \ln[\sinh(\sigma)/\sigma] \\
f_c \ &=& \ [1/\alpha] 
\eeq 
Again we emphasize that those sharp transitions appear only in the limit ${L\rightarrow \infty}$.

\begin{figure}[b!]
\centering 
(a) 
\includegraphics[width=0.95\hsize]{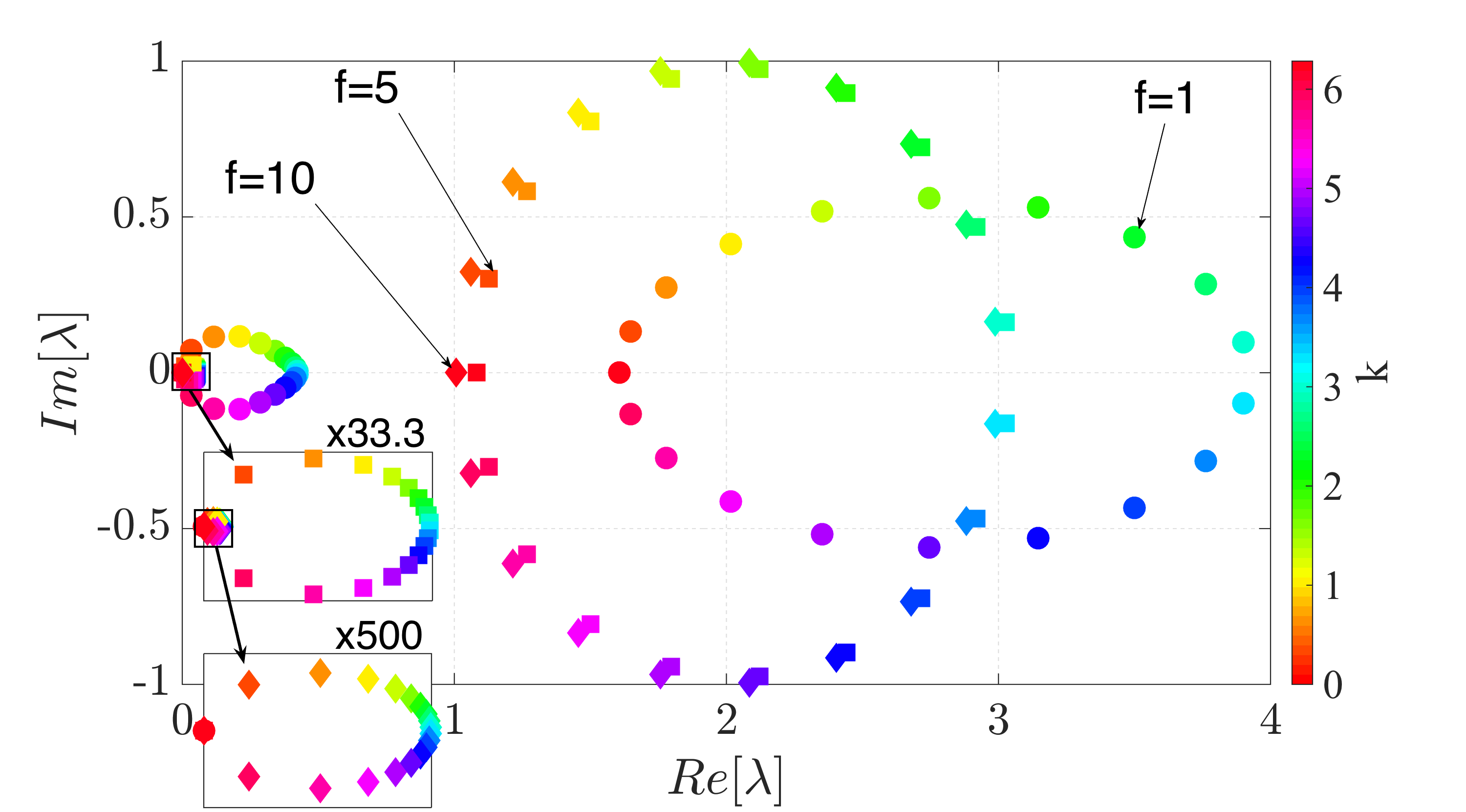}  

(b) 
\includegraphics[width=0.95\hsize]{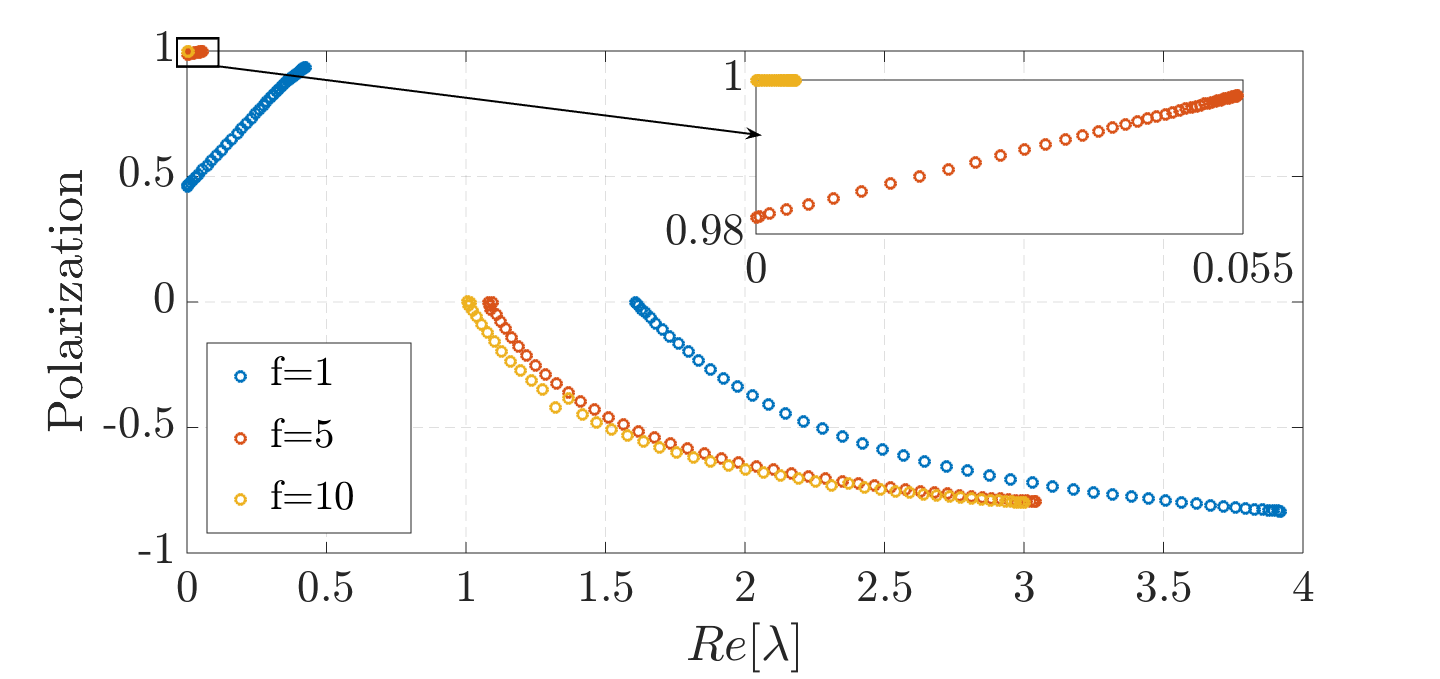} 
\caption{ 
{\bf The relaxation spectrum for a non-disordered ring with dangling bonds.}
In panel (a) we display in the complex plane the eigenvalues $\lambda_{\nu}(k)$ for a system of length $L=100$. 
Here and in the next figures ${w_o=c_o=1}$ and ${\alpha=1/2}$.
There are 3 spectra that correspond to $f=1$ (circles), and $f=5$ (squares), and $f=10$ (diamonds).
The points are color-coded by~$k$. 
Panel (b) displays the polarization of the associated eigenmodes.}
\label{f5} 
%
%
\centering 
\includegraphics[width=\hsize]{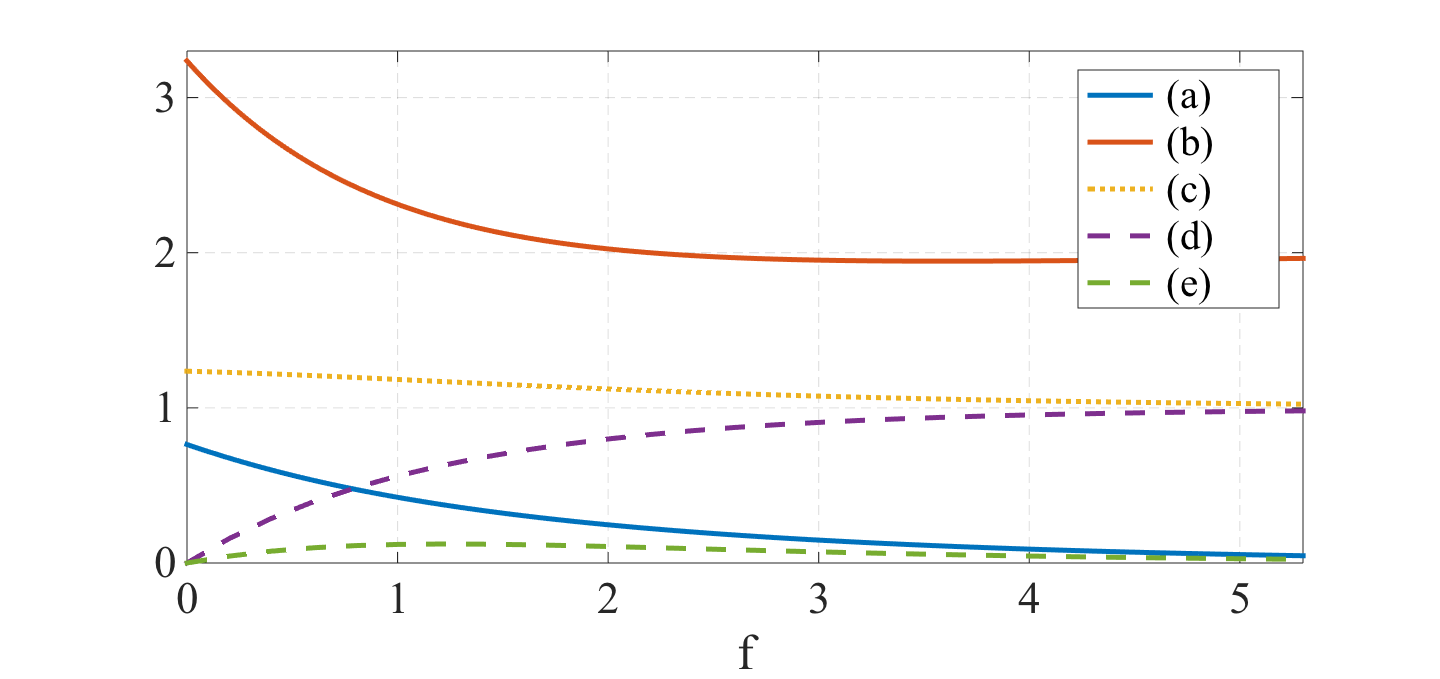}  
\caption{ 
{\bf The band boundaries of the spectrum versus bias.}    
The boundaries of the two bands that are displayed in \Fig{f5} are plotted versus~$f$: 
(a)~$\lambda_0(\pi)$; 
(b)~$\lambda_1(\pi)-\lambda_1(0)$; 
(c)~$\lambda_1(0)-\lambda_0(\pi)$;
(d)~$\max[\im(\lambda_{\nu=1})]$;
(e)~$\max[\im(\lambda_{\nu=0})]$.
The $\nu=0$ band is responsible for the long-time under-damped relaxation.  
It becomes tiny for large bias, implying a longer relaxation time.  
}
\label{f6}
\end{figure}

\section{Relaxation spectrum for a non-disordered ring} 

Up to now we have focussed on the NESS, which is the $\lambda{=}0$ eigenstate of $\bm{\mathcal{W}}$. Now we turn to discuss the whole spectrum. From the full spectrum we can derive results not only for the drift velocity, but also for the diffusion coefficient. Furthermore, it is the spectrum that contains the information on the delocalization transition, and in particular on whether the relaxation is over-damped or under-damped. We first address the non-disordered version of \Fig{f1}b.

The relaxation modes are the right eigenvectors of $\bm{\mathcal{W}}$, 
and they satisfy the equation $\bm{\mathcal{W}} \Psi = -\lambda \Psi$. 
In the absence of disorder, due to Bloch theorem, 
the matrix becomes block diagonal (see Appendix~A).
Consequently the eigenvalues are labelled as $\lambda_{\nu}(k)$, 
where $k$ is the wavenumber (the Bloch phase), and ${\nu=0,1}$ is the band index. 
For each~$k$ we have to diagonalize a $2\times 2$ matrix:
\beq 
\bm{\mathcal{W}}^{(k)} = 
\left(\amatrix{
(e^{-ik}{-}1)w^{+} + (e^{ik}{-}1)w^{-} -c^{+}  & c^{-} \cr  c^{+} & -c^{-}
}\right) \ \ \ \ \ 
\eeq
The result of the diagonalization is demonstrated graphically in \Fig{f5}. 
The NESS is the eigenstate that is associated with  ${\lambda_{0,0}=0}$.
The two bands of the spectrum form two complex bubbles in the complex $\lambda$ plane.
The points are color-coded by the Bloch phase~$k$.
The second panel in  \Fig{f5} provides further information on the polarization of the eigenmodes, 
which is defined as ${D=|\Psi_{\downarrow}|^2-|\Psi_{\uparrow}|^2}$, 
where the eigen-vector ${\Psi \mapsto (\Psi_{\uparrow},\Psi_{\downarrow}) }$ of $\bm{\mathcal{W}}^{(k)}$ assumes standard normalization.
Note that the two bands have opposite polarity. In particular note that the NESS 
is polarized ``positively", reflecting that the bias expels the probability 
from the chain and push it into the dangling sites.

The boundaries of the two bands are displayed in \Fig{f6}, 
and are based on the expressions that can be found in Appendix~B. 
Our interest is mainly in the $\nu=0$ band that is bounded from below by $\lambda=0$. 
Its complexity implies under-damped relaxation in the long time limit. 
For large bias it becomes tiny, implying a longer relaxation time.

The weighted drift velocity can be calculated from the spectrum \cite{neq}.   
The result is in agreement with \Eq{e4}, indicating that the Math is self-consistent: 
different ways lead to the same expression. Namely, 
\beq
v = i\frac{\partial \lambda_0(k)}{\partial k}\bigg|_{k=0} 
\ = \  w_o \left( \frac{1-e^{-f}}{1+e^{\alpha f}} \right)
\eeq
The diffusion coefficient can be calculated as well:  
\beq \label{eD}
&& \mathcal{D} \ \ = \ \ \frac{1}{2}\frac{\partial^2 \lambda_0(k)}{\partial k^2}\bigg|_{k=0} 
\\ \nonumber
&& \ =  
\left[\frac{1+e^{-f}}{1+e^{\alpha f}} \right]\frac{w_o}{2}
+ \left[\frac{(1-e^{-f})^2}{(1+e^{\alpha f})^3} e^{2\alpha f} \right]\frac{w_o^2}{c_o}
\eeq
\Fig{f2} illustrates the $f$ dependence of the above expressions that were derived 
for the network of \Fig{f1}b, and also the result (see Appendix~B) that applies to the network of \Fig{f1}c.
In a simulation we can visualize the evolving probability distribution 
as a stretching cloud. The second term in \Eq{eD}, that diverges in the ${c_o\rightarrow 0}$ limit, reflects the departure of a drifting piece along the chain, 
from remnants that lags in the dangling bonds.
The first term reflects the extra spreading of the drifting piece. 
In the zero bias limit (${f\rightarrow 0}$) it is only the latter contribution 
that survives, leading to ${\mathcal{D} \rightarrow 1/2}$.

\section{Relaxation spectrum for a disordered ring} 

\begin{figure}[b!] 
\centering 
(a)~\includegraphics[width=0.9\hsize]{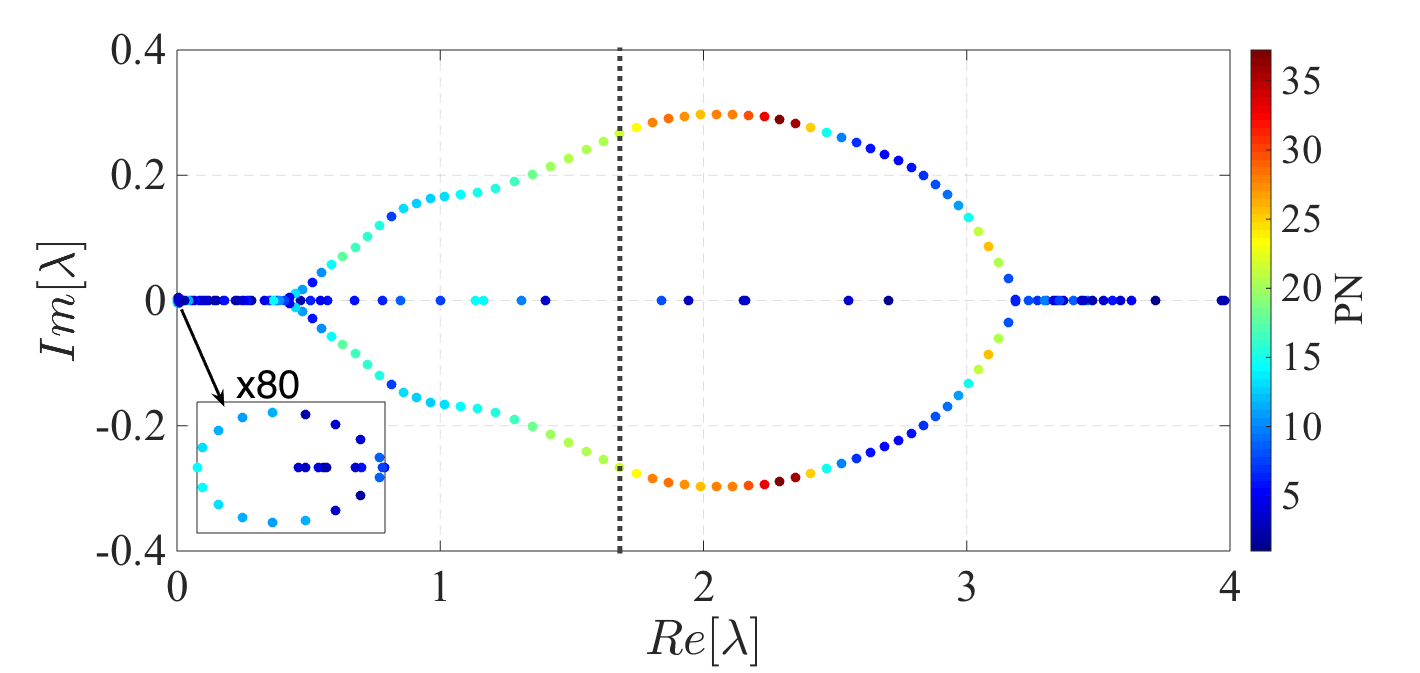}
(b)~\includegraphics[width=0.9\hsize]{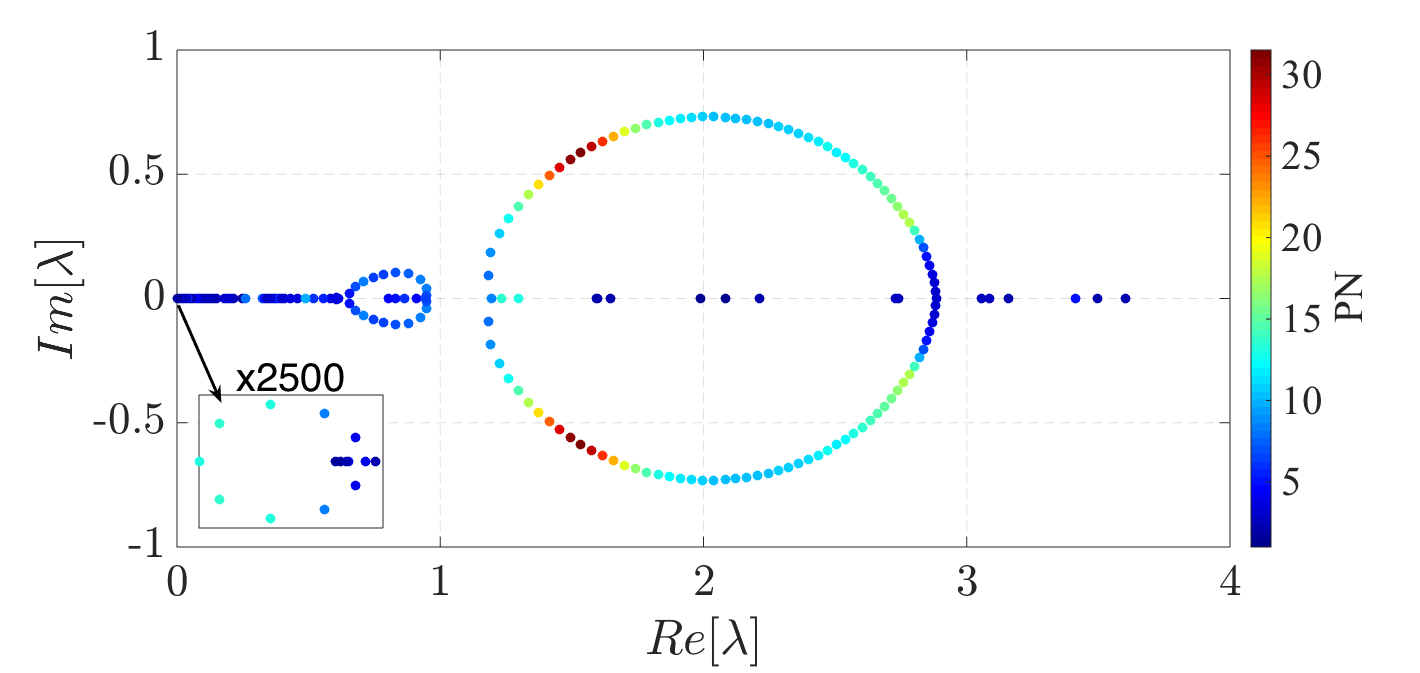} 
(c)~\includegraphics[width=0.9\hsize]{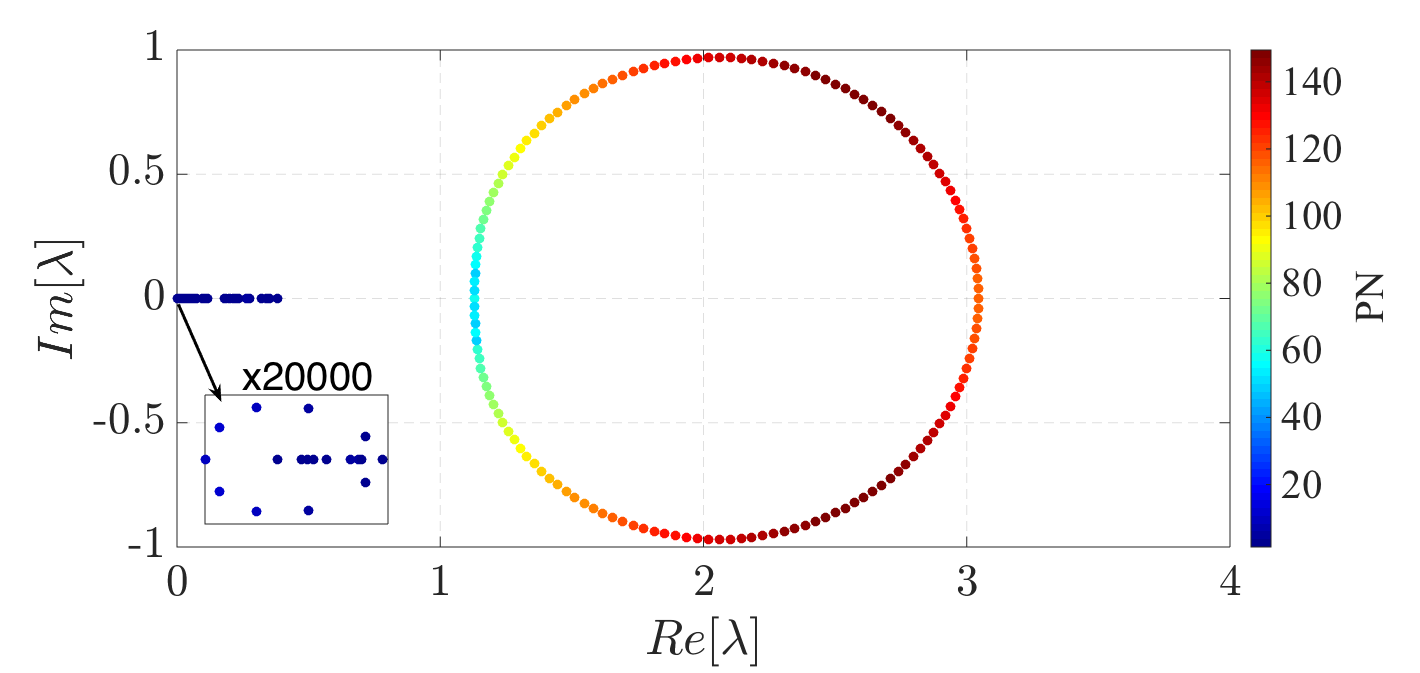}  
(d)~\includegraphics[width=0.9\hsize]{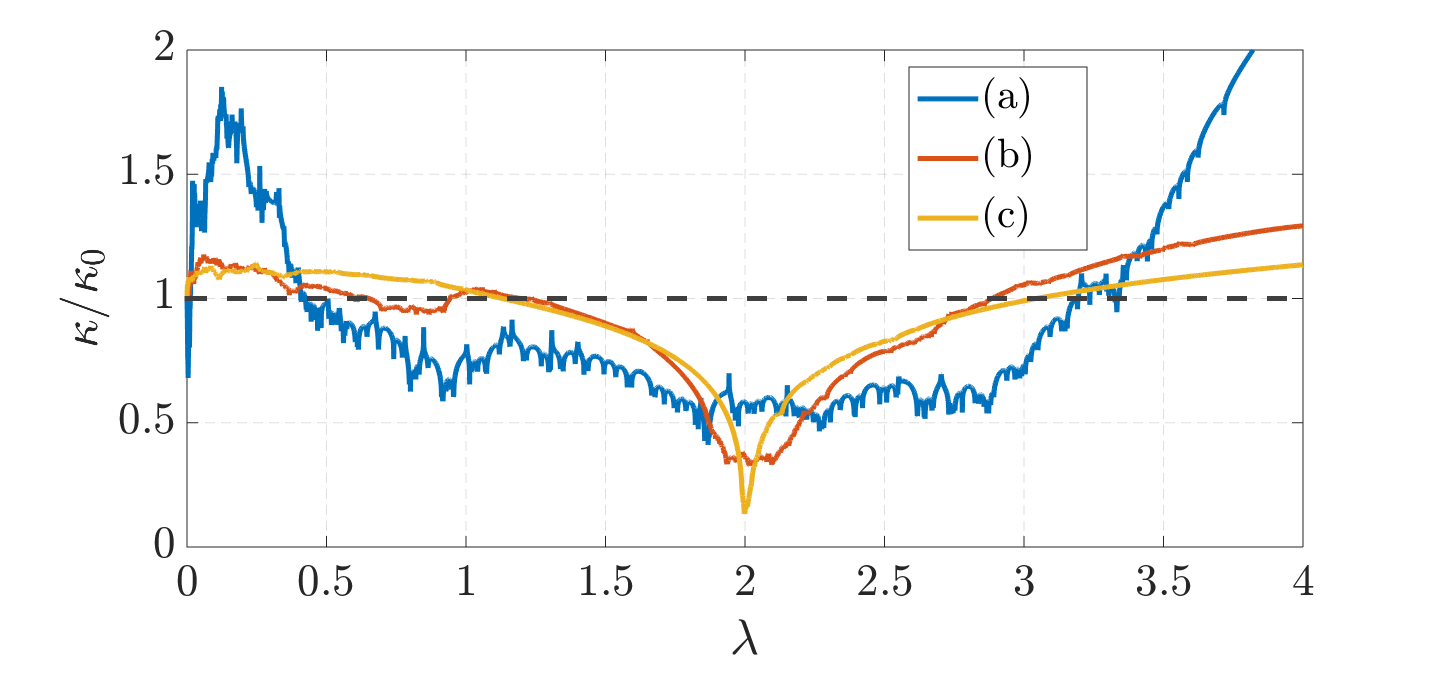} 
\caption{
{\bf The relaxation spectrum for a disordered ring with dangling bonds.} 
We display in the complex plane the eigenvalues for a system of length $L{=}150$. 
The other parameters are the same as for the ring of \Fig{f5}, 
with added off-chain disorder of strength ${\sigma{=}5}$. 
Panels (a-b-c) are for ${f=1,5,10}$ respectively. 
The points are color-coded by the participation number (PN), 
namely, the number of units cells that are occupied by the eigenmodes. 
The vertical dotted line in panel~(a) is a median 
that divides the spectrum into two equal groups. 
The spectrum separates into two bands in panels~(b,c).
Panel~(d) is related to the analysis in Section~VII.    
The inverse localization length is defined by \Eq{e39}, 
and calculated from the $\bm{\mathcal{W}}$ of panels (a-b-c).  
Complex roots are expected if $\kappa(\lambda)<\kappa(0)$.
}
\label{f7}
\end{figure}

\begin{figure} 
\centering 
\includegraphics[width=8cm]{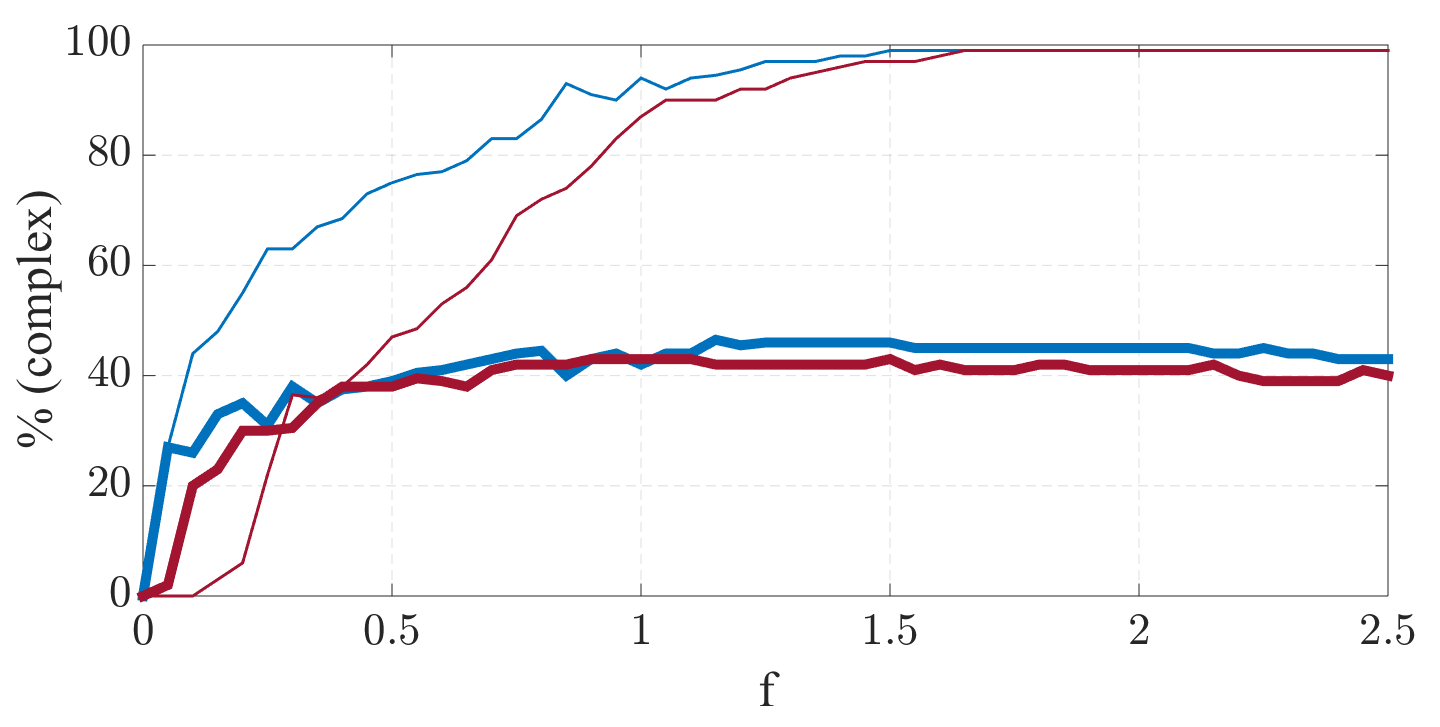}  
\includegraphics[width=8cm]{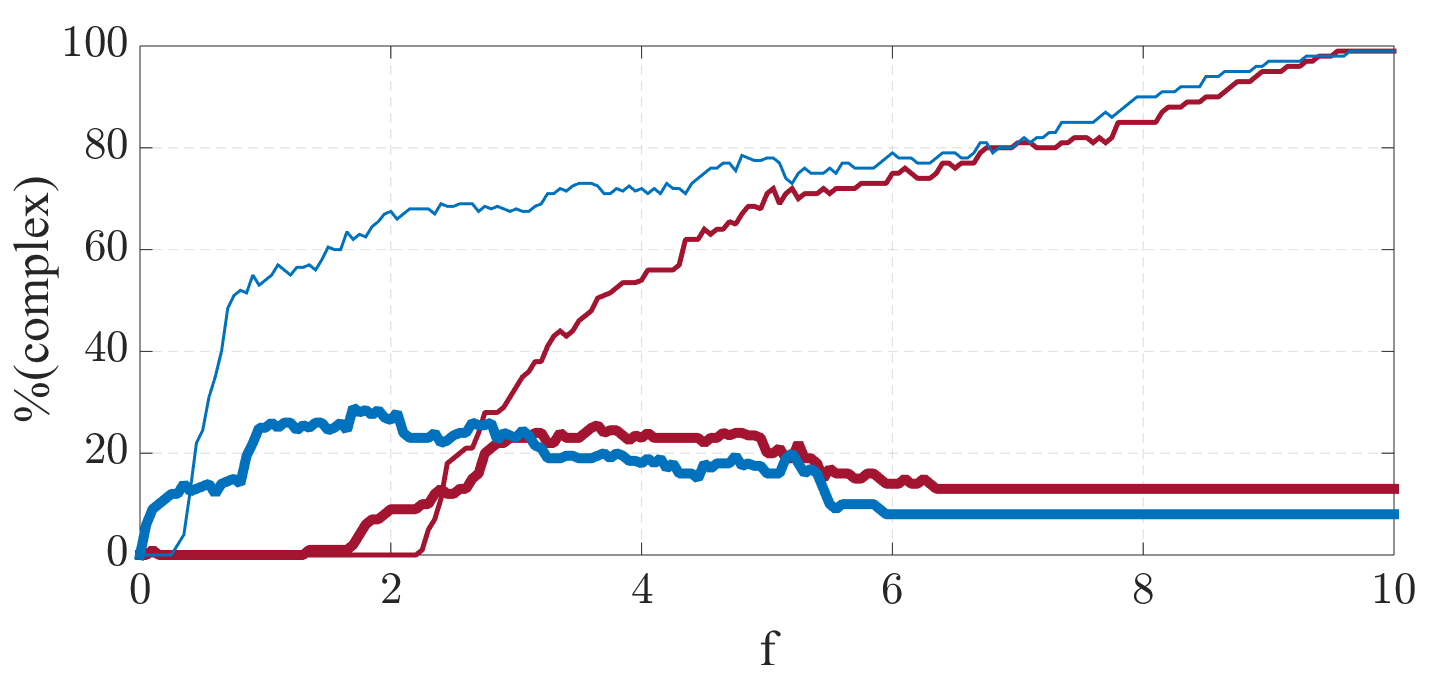} 
\caption{
{\bf Delocalization of the eigenstates.}
The fraction of complex eigenvalues (indicating delocalized eigenstates)
is calculated separately for each band (thicker lines for ${\nu=0}$).
If the bands are not separated we use median for their practical definition.
The upper and lower panels are for disorder of strength $\sigma=1,5$ respectively.
Both off-chain (blue) and on-chain (red) disorder are considered. 
For on-chain disorder the delocalization transition is clearly resolved for ${\sigma=5}$, 
while for off-chain disorder the ${\nu=0}$ band always feature a complex bubble.
For strong bias the  ${\nu=0}$ band exhibits complexity saturation, 
while the  ${\nu=1}$ band becomes 100\% complex irrespective of the disorder type.
}
\label{f10}
\end{figure}

In the absence of disorder the spectrum of a biased system features a ``complex bubble" that touches the origin, indicating finite drift velocity and under-damped relaxation. Adding disorder some of the eigenmodes become localized, and the associated eigenvalues become real. If the ``complex bubble" is diminished near the origin, over-damped relaxation is implied. 

In the model of \Fig{f1}b the forward and backward transition rates $w_n^{\pm}$ are along 
the $n$th bond that connects the $n$ and $n-1$ sites,  
while the $c_n^{\pm}$ are for the forward and backward rates along the dangling bonds. Namely,   
\beq
w_{n}^{+} &=& w_o \\ 
w_{n}^{-} &=& w_o e^{-f_n} \\
c_{n}^{+} &=& c_o \\
c_{n}^{-} &=& c_o e^{-g_n} 
\eeq
We assume that the $f_n$ contain random box-distributed 
on-chain disorder due to environmental irregularities, as defined in \Eq{eOCD}, 
while the ${g_n = \alpha f + \text{random}[-\sigma,\sigma]}$ 
reflect off-chain disorder with the same distribution.  
We use ${\alpha=1/2}$ in the subsequent numerics.   
Representative results for the spectrum are presented in \Fig{f7}. 
Each point is color-coded by the participation number (PN), 
namely, the number of units cells that are occupied by the associated eigenmode. 
With standard normalization the definition is 
\beq
\text{PN} = \left( \sum_n Q^2_n \right)^{-1}; 
\ \ \ \ Q_n = |\psi_{n\uparrow}|^2+|\psi_{n\downarrow}|^2
\eeq
We see clearly that large PN is correlated with complexity. 
This is expected from the general phenomenology of the delocalization transition \cite{nelson3}, 
namely, a localized eigenstate is effectively living on a disconnected ring, for which the asymmetry of the transition can be gauged-away (the technical aspect will become clear in the next section, where we discuss the secular equation for the eigenvalues).  
We note that the average polarization of the eigenmodes (numerical results not displayed) 
is similar to that of a non-disordered case (see \Fig{f5}b).   


\Fig{f10} displays the fraction of complex eigenvalues.
The fraction is calculated separately for each band.
If the bands are not separated by a gap, we use median for their practical definition.
Namely, 50\% of the eigenvalues that have the lowest $\re[\lambda]$ are defined 
as the ${\nu=0}$ band, bounded from above by a vertical dotted line in \Fig{f7}a.
For on-chain disorder the delocalization transition is clearly resolved if the disorder 
is strong enough. Namely, up to some critical value of $f$ the complex bubble that touches the origin 
disappears, and the eigenvalues there become real. 
Complex eigenvalues with large $\re[\lambda]$ may exist: they represent a transient under-damped relaxation. 
For long times the predominant dynamical behavior is over-damped 
if the bias is below the delocalization threshold.  
We see that such delocalization transition does not appear if we have only off-chain disorder.
In the latter case a small complex bubble that touches the origin survives even if the disorder is large, 
irrespective of the bias.         
For strong bias the  ${\nu=0}$ band exhibits complexity saturation,
meaning that a finite fraction of real eigenvalues survives. 
This complexity saturation will be explained by reduction to the standard case (see next section).  
In contrast the ${\nu=1}$ band becomes 100\% complex irrespective of the disorder type.

\section{The localization of the eigen-modes} 

In order to analyse the delocalization that we observe in \Fig{f10} for the model of \Fig{f1}b,  
we show that its characteristic equation, ${\det(\lambda+\bm{\mathcal{W}})=0}$, 
can be reduced to that of an effective Sinai model (\Fig{f1}a).
Then we can follow the same strategy as in \cite{psl,nelson1,nelson2,nelson3} 
that relates the (possibly complex) spectrum of the non-hermitian matrix $\bm{\mathcal{W}}$ 
to the real spectrum of an associated hermitian matrix $\bm{H}$. 
We note that a localized eigenmode is effectively living on a disconnected ring. 
For a disconnected ring the spectrum of  $\bm{\mathcal{W}}$ is identical 
to the spectrum of $-\bm{H}$ and therefore has to be real. This is the reason   
for associating the term ``delocalization" with the complexity of the spectrum.

\begin{figure} 
\centering 
(a)~\includegraphics[width=0.9\hsize]{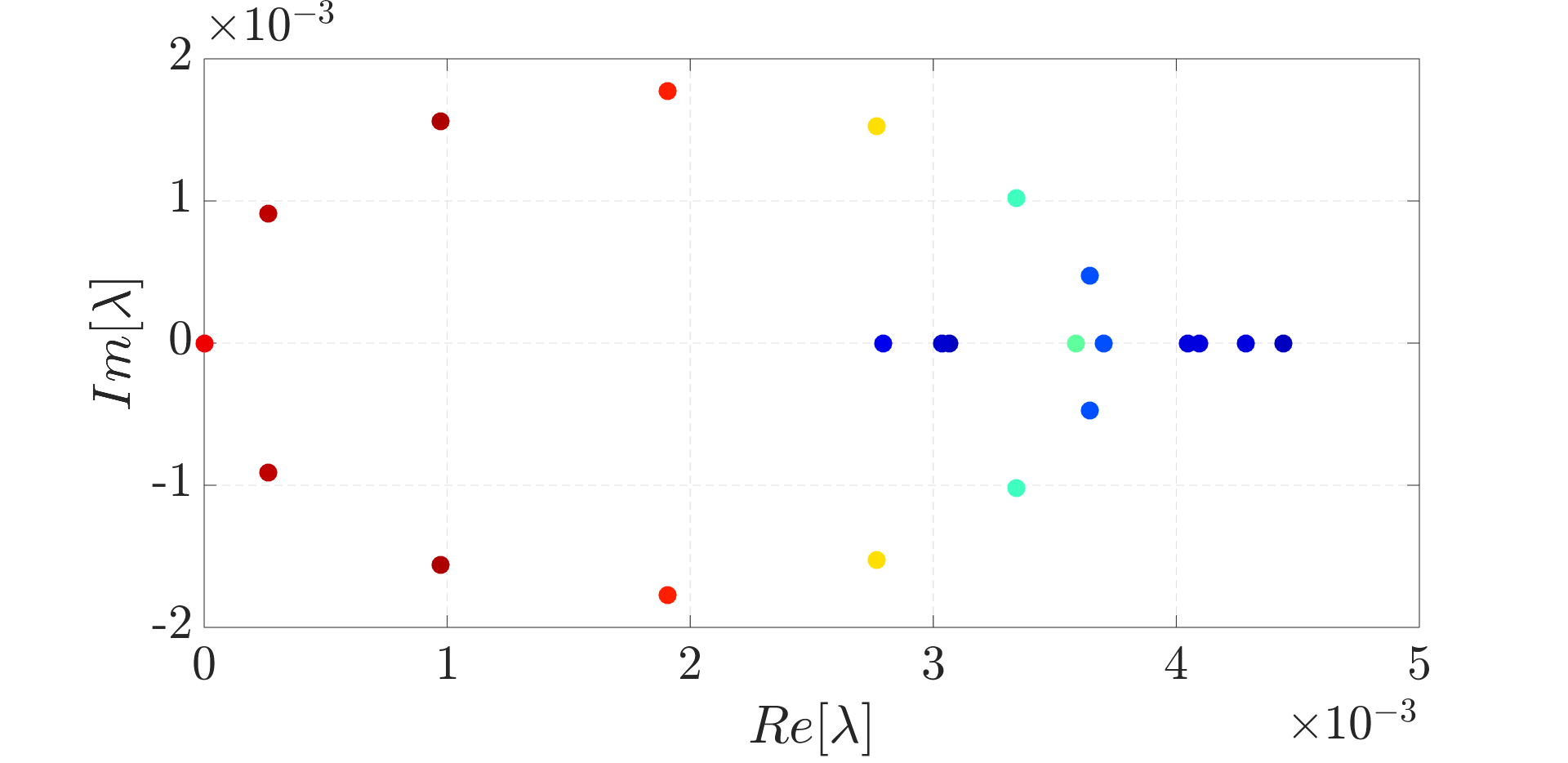} 
(b)~\includegraphics[width=0.9\hsize]{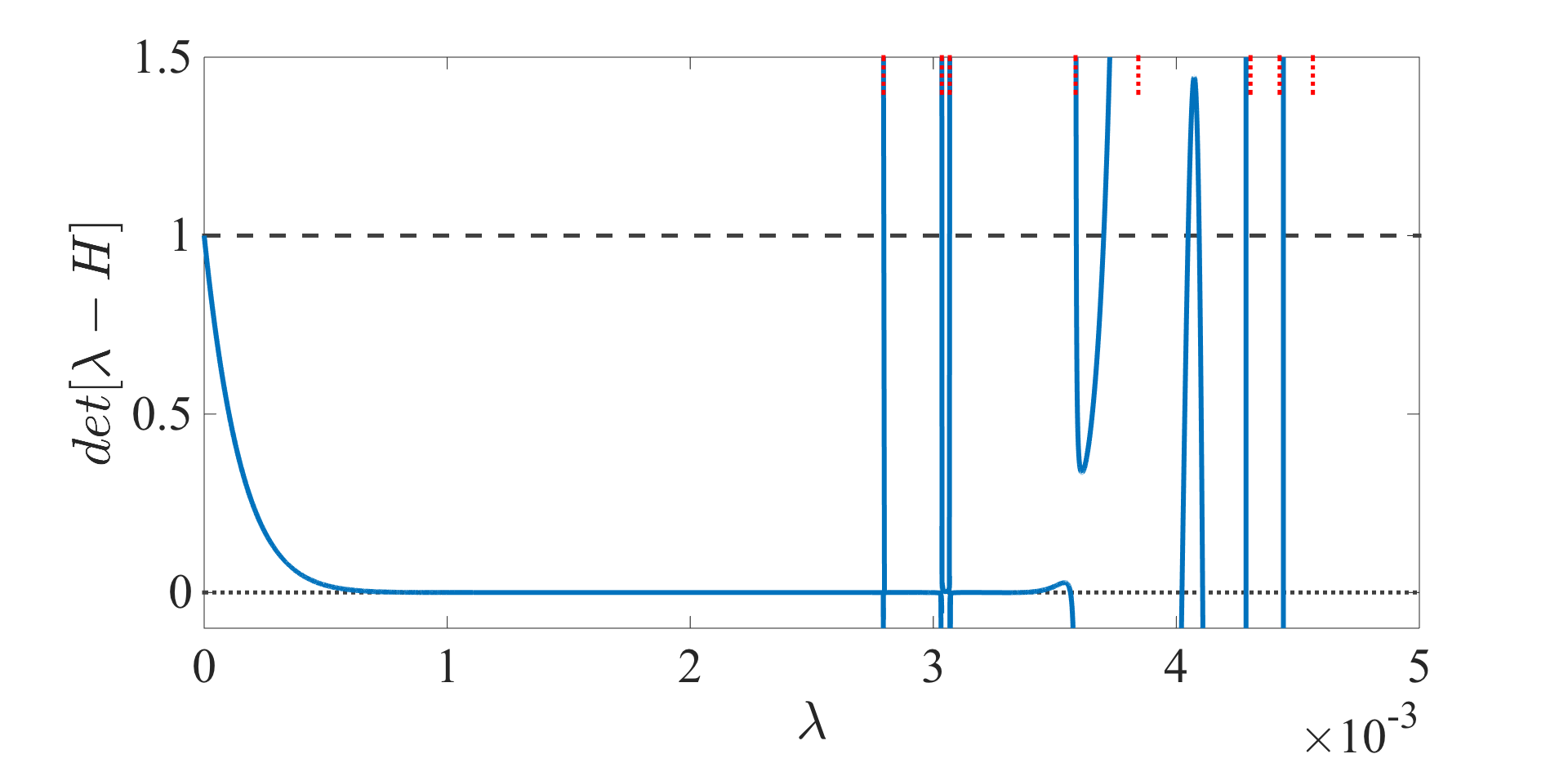} 
(c)~\includegraphics[width=0.9\hsize]{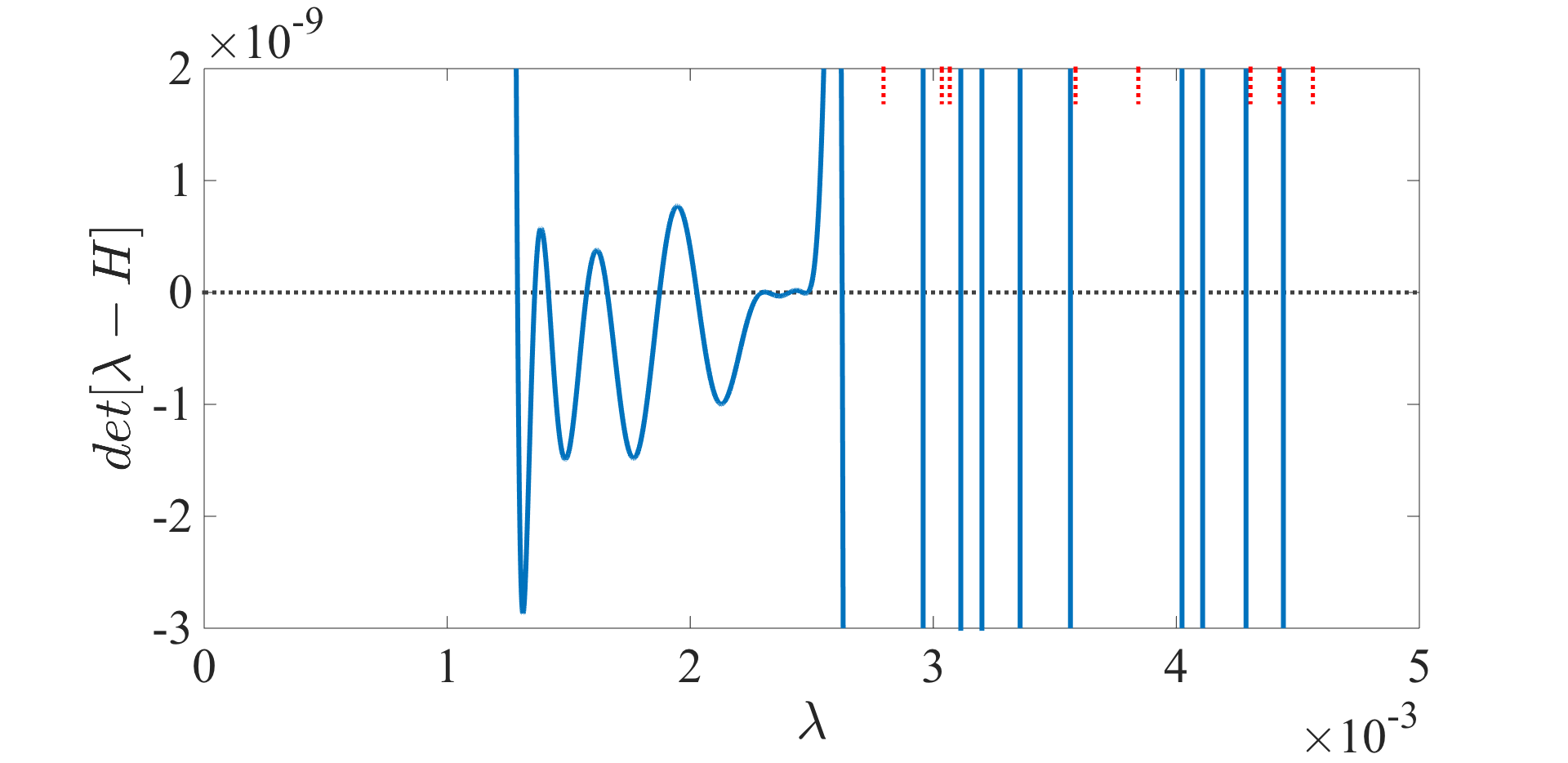} 
\caption{ 
{\bf Spectral determinant for off-chain disorder.} 
We consider an $L{=}150$ system with off-chain disorder $\sigma{=}5$ and $f{=}2$.
Panel~(a) displays a the relaxation spectrum, 
focusing in the ${\lambda \sim 0}$ region where complex and real eigenvalues coexist.
Panel~(b) displays the spectral determinant ${\det(\lambda-\bm{H})}$ 
along the real axis. Each intersection with the dashed line 
implies a real eigenvalue of $\bm{\mathcal{W}}$ (see text for further details). 
The upper bars indicate the values ${\lambda=c_n^{-}}$ at which the 
determinant diverges. Panel~(c) provides a vertical zoom.  
} 
\label{f9s}
\end{figure}

\sect{The Reduction}
The equation $\bm{\mathcal{W}}\psi =-\lambda \psi$ 
for the eigenmodes of a chain with dangling bonds is 
\beq \nonumber
w_{n}^{+} \psi_{n{-}1} + w_{n+1}^{-}\psi_{n{+}1} + c_{n}^{-} \psi_{n,\downarrow}-\gamma_n\psi_{n} &=& -\lambda\psi_{n} 
\\ \nonumber
c_{n}^{+}\psi_{n} - c_{n}^{-}\psi_{n,\downarrow} &=& -\lambda\psi_{n,\downarrow}
\eeq
where ${\gamma_n =  w_{n+1}^{+} + w_{n}^{-} + c_{n}^{+}}$, 
and we have used the simplified notation $\psi_n \equiv \psi_{n,\uparrow}$. 
Eliminating the dangling bonds from the set of coupled equations 
we get the single-channel tight binding equation
\beq \label{e33}
w_{n}^{+} \psi_{n{-}1} + w_{n+1}^{-} \psi_{n{+}1} - u_n(\lambda) \psi_n = -\lambda \psi_n 
\eeq 
with 
\beq \label{e34}
u_n(\lambda) \ &\equiv& \ [w_{n{+}1}^{+} + w_{n}^{-} + c_{n}^{+}] - \frac{c_{n}^{+} c_{n}^{-}}{c_{n}^{-} - \lambda}
\\ \ & \approx &  (1+e^{-f_n}) w_o - \lambda \, e^{g_n} 
\eeq
The second line above is a small $\lambda$ expansion. 
We see that off-chain disorder (random $g_n$) introduces diagonal disorder of intensity that is proportional to $\lambda^2$, 
while on-chain disorder (random $f_n$) does not vanish in this limit.   
This already explains qualitatively why the $\lambda$ spectrum is hardly affected 
by off-chain disorder in the vicinity of the origin.

\sect{The Spectral determinant}
We proceed with a quantitative treatment of the characteristic equation ${\det(\lambda+\bm{\mathcal{W}})=0}$. After the reduction to a single-channel tight binding model it takes the form ${\det(\lambda + \bm{W})=0}$. The ${N \times N}$ matrix $\bm{W}$ (note the different font), with ${N=L}$, is associated with the reduced equation, namely, \Eq{e33}. For the calculation of the determinant it is convenient to write the forward and backward rates as $w_n \exp(\pm f_n/2)$, where 
\beq \label{ewn}
w_n  \ \equiv \  \sqrt{w_n^{+}w_n^{-}}  \ = \ w_o \exp(-f_n/2)
\eeq
The affinity is defined as $\sum_n f_n = Nf$, where $f$ is what we called "bias".  
%
%
Then it is possible to define an associated hermitian matrix $-\bm{H}$ that has the same 
diagonal elements, while the  off-diagonal couplings are $w_n$. 
A linear-algebra formula \cite{blockmatrix} (optionally see Appendix~C of \cite{rss}) 
leads to the identity 
\beq
&& \det(\lambda + \bm{W}) = \det(\lambda - \bm{H}) \nonumber \\ 
&& \ \ \ - \left[e^{Nf/2} + e^{-Nf/2} - 2\right] \prod_n (-w_n)
\eeq

\sect{Finding the eigenvalues}
It is convenient  to define the average coupling as   
\beq
w_{\text{avg}} \ &=& \ \left[\prod_{n=0}^{N} w_n\right]^{1/N} 
\eeq
Then the characteristic equation takes the form:
\beq \label{eT}
\prod_{k=0}^{N}\left(\frac{\lambda - \epsilon_k(\lambda;f)}{-{w_{\text{avg}}}} \right) \ &=& \ 2\left[\cosh\left(\frac{Nf}{2}\right) - 1\right] 
\eeq
where $\epsilon_k$ are the real eigenvalues of the hermitian $\bm{H}$ matrix.
A numerical demonstration of this ({\em exact}) equation is provided in \Fig{f9s}. 
Its left hand side, up to  factor, is the spectral determinant $Z(\lambda) = \det(\lambda - \bm{H})$. 
The right hand side is a constant $Z_0$ that is represented by a dashed horizontal line.
Note that the dashed line intersects the spectral determinant at ${\lambda=0}$, 
which corresponds to the NESS, hence ${Z_0=Z(0)}$. The non-trivial {\em real} eigenvalues 
of $\bm{\mathcal{W}}$ are determined by the intersection of $Z(\lambda)$ with the dashed line.

In the absence of dangling bonds $Z(\lambda)$ is a polynomial with 
roots $\epsilon_k$ that do not depend on $\lambda$. In such case it oscillates 
around zero with some {\em envelope}. Taking the log of this envelope we define 
a function $\kappa(\lambda)$ such that 
\beq
|Z(\lambda)| \ \ \lesssim \ \ w_{\text{avg}}^N  \ e^{N \kappa(\lambda)} 
\eeq     
In \Fig{f9s}b, up to ${\lambda \sim 2.6}$ the envelope of $Z(\lambda)$ 
is well below the dashed line, and consequently all the eigenvalues in this 
region are complex. More generally there might be regions where the envelope  
is above the dashed line, and then we get real eigenvalues, as demonstrated in \Fig{f7}.

Due to the elimination of the dangling bonds the $\epsilon_k$ 
acquire $\lambda$ dependence: for each $\lambda$ we have to calculate 
again the $\epsilon_k(\lambda;f)$ spectrum. The most conspicuous implication 
is the appearance of singular spikes at ${ \lambda =  c_n^{-} }$. 
For large enough~$f$ those spikes invade the lower band of the spectrum, 
as demonstrated in \Fig{f9s}b. Extra real eigenvalues that co-exist 
with the complex bubble are implied. Those real eigenvalues corresponds 
to over-damped relaxation modes that are localized on the dangling sites.

\sect{The Thouless relation}
Following \cite{nelson3} we point out that the log of the left hand side in \Eq{eT}, 
after dividing by $N$, is the Thouless formula \cite{Thouless,ThoulessR} 
for the inverse localization length $\kappa$ in the hermitian problem. 
Substitution of the definition of $w_n$ leads to 
\beq \label{e39}
\kappa(\lambda) \ \ = \ \ \frac{f}{2} \ + \ \frac{1}{N}\sum_{k=0}^{N} \ln \left|\frac{\lambda - \epsilon_k(\lambda;f)}{w_o} \right|
\eeq
A few words are in order regarding the use of this formula. 
It is implicit that we refer here to the {\em envelope} of the spectral determinant. 
Furthermore, the identification  of $\kappa(\lambda)$ as the inverse localization length 
is meaningful only in the $\lambda$~range where the (real) spectrum stretches, 
otherwise it is a merely a formal continued expression.

\begin{figure} 
\centering 
\includegraphics[width=0.9\hsize]{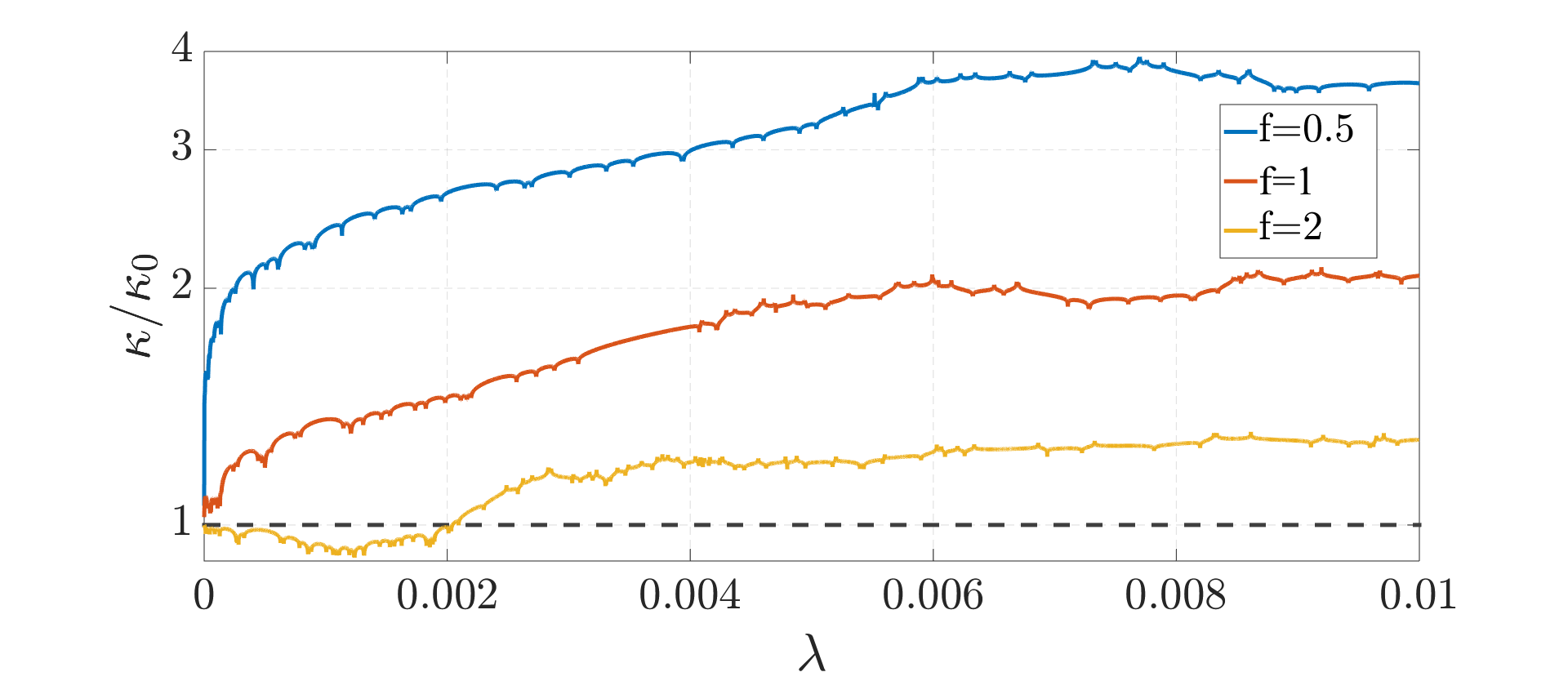}  
\includegraphics[width=0.9\hsize]{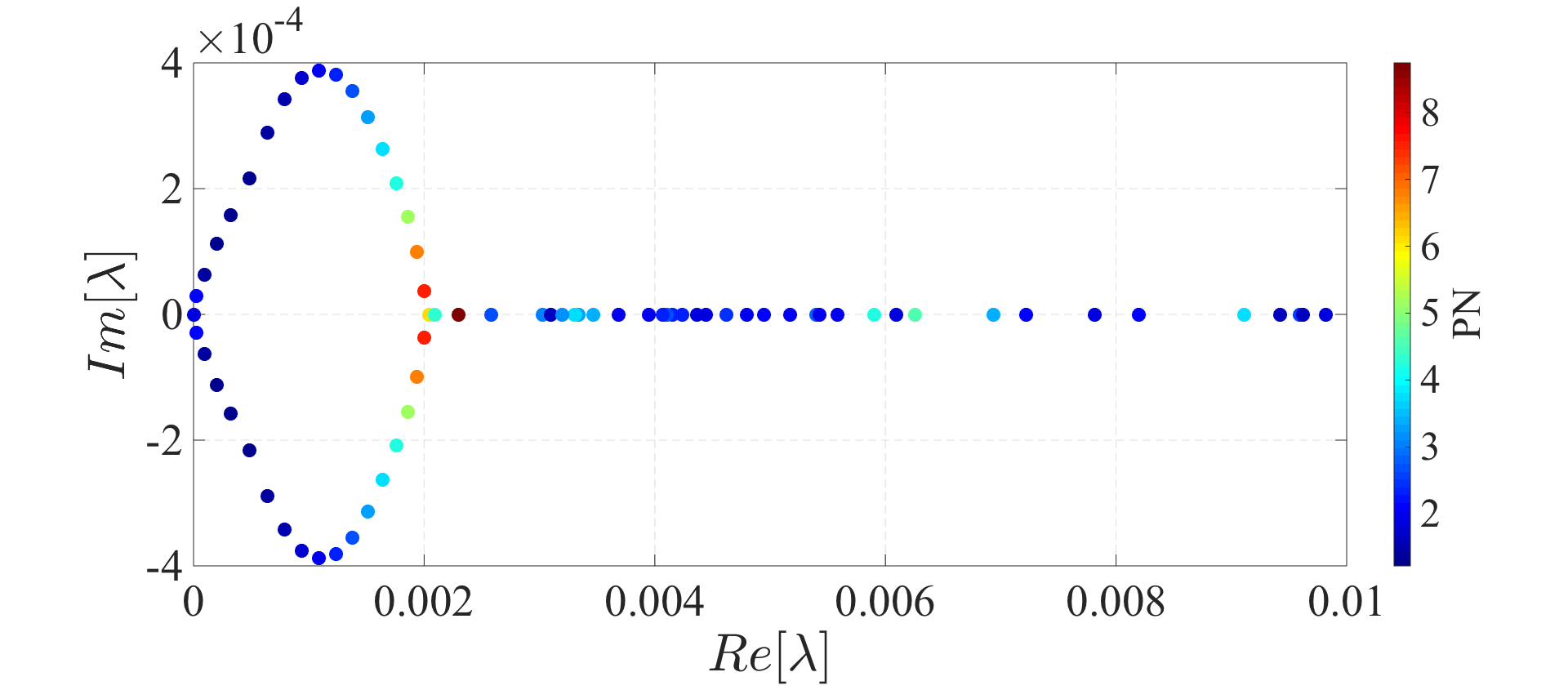} 
\caption{ 
{\bf The inverse localization length for on-chain disorder.} 
We consider an $L=300$ chain with both off-chain and on-chain disorder ${\sigma=5}$.
The inverse localization length $\kappa(\lambda)$ of \Eq{e39}
is calculated in the $\lambda$ range of the ${\nu=0}$ band.
The bias ${f=0.5,1,2}$ is indicted in the legend.
In the lower panel we display the spectrum for ${f=2}$.
The spectrum is complex in the range where ${\kappa(\lambda)<\kappa(0)}$. 
}
\label{f8}
\end{figure}

\begin{figure} 
\centering 
\includegraphics[width=0.9\hsize]{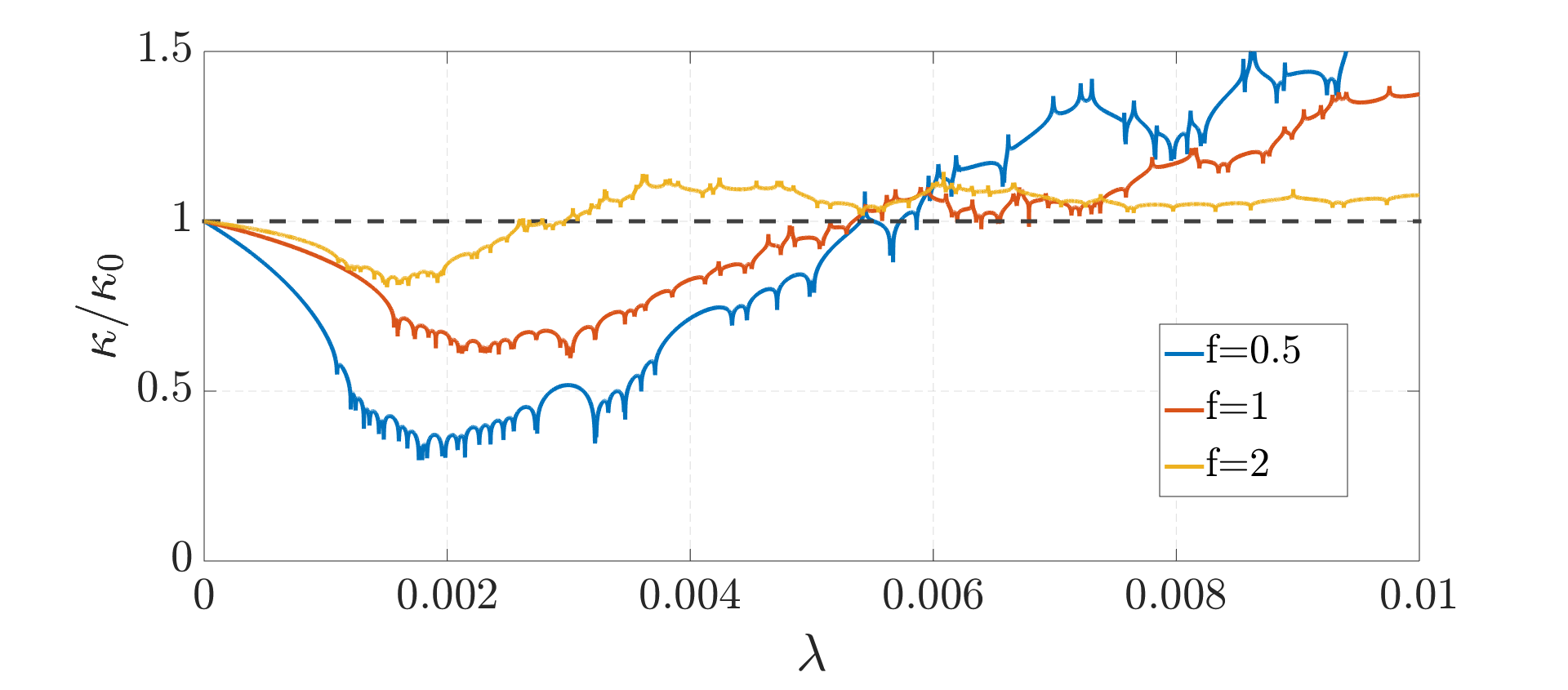} 
\includegraphics[width=0.9\hsize]{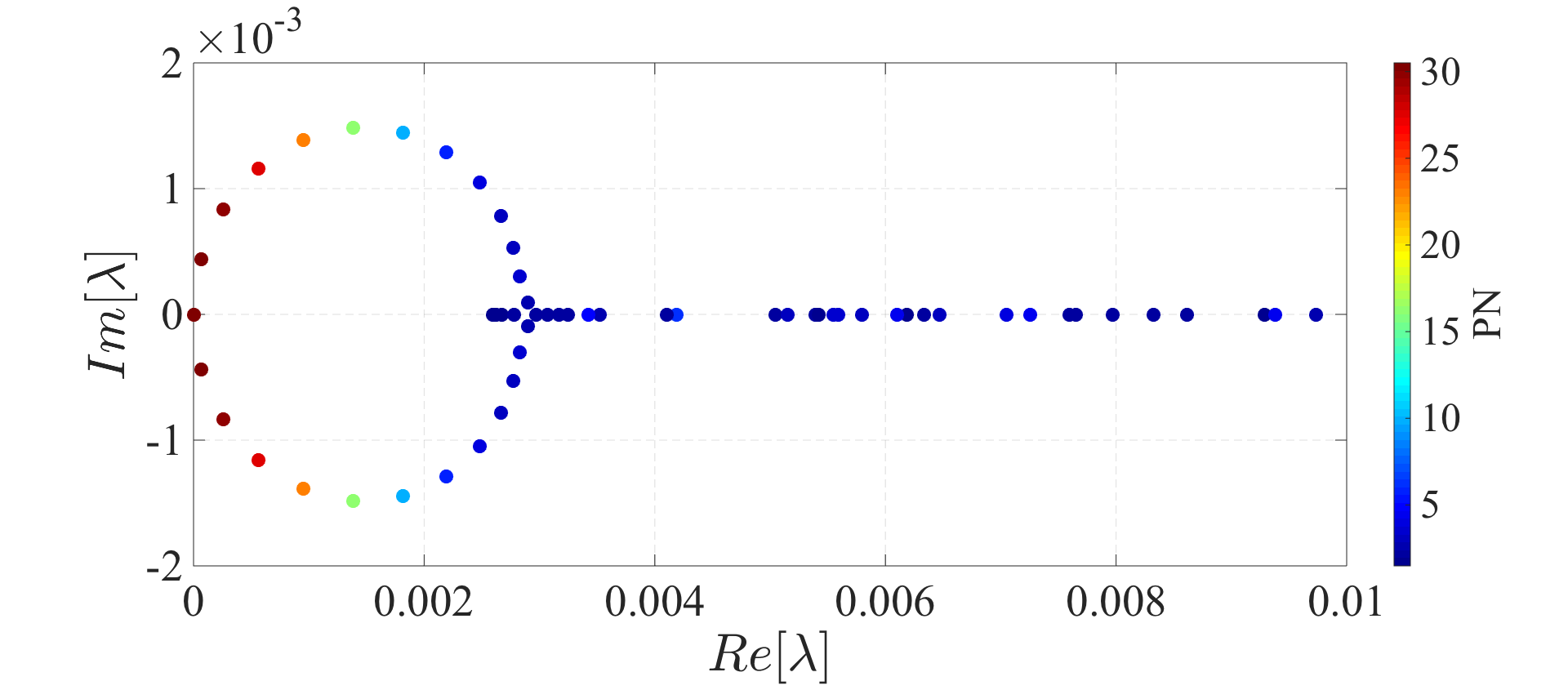} 
\caption{ 
{\bf Inverse localization length for off-chain disorder.} 
The same as for \Fig{f8}, but with only off-chain disorder.
Here the spectrum has a complex fraction for any $f$.
}
\label{f9}
\end{figure}

Traditionally $\kappa$ is determined using a transfer matrix method.
There are also some analytical approximations that can be used (see next paragraph).  
But for our purpose a direct numerical calculation using the Thouless formula is most convenient. 
The reduction of the $\bm{\mathcal{W}}$-problem to the $\bm{H}$-problem assumes real $\lambda$, 
hence \Eq{eT} provides real roots if ${\kappa(\lambda) > \kappa(0)}$, 
otherwise complex  spectrum should appear (which cannot be extracted directly by inspection). 
This expectation is confirmed numerically by \Fig{f8} and \Fig{f9}, 
where we contrast the delocalization scenario for on-chain disorder 
with the scenario that is observed for off-chain disorder. 
In the latter case there is always a small range near the origin where ${\kappa(\lambda) < \kappa(0)}$, 
leading to the appearance of a complex bubble that implies under-damped relaxation.

\sect{The inverse localization length}
As mentioned above, there are some analytical approximations that can be used in order to evaluate $\kappa$. Exact results are available in the continuum limit, which is not useful here. We are therefore satisfied with a standard Born-approximation that is based on a Fermi-Golden-Rule picture: 
\beq \label{eLOC}
\kappa(\lambda) \ = \ \frac{\sigma_{\parallel}^2}{8 \bar{w}^2 k_{\lambda}^2} 
\ + \  \frac{\sigma_{\perp}^2}{8 \bar{w}^2}k_{\lambda}^2  
\eeq  
Those are Eqs(14-15) of \cite{rna}, where further refinements are discussed, and additional references therein. This equation requires a careful explanation. It is expressed in terms of the wavenumber that is determined by the dispersion relation 
${\lambda = 2\bar{w} (1-\cos(k)) }$, where $\bar{w}$ is the harmonic average over the bond couplings \Eq{ewn}.  The approximation ${\lambda \approx w_o k^2 }$ can be used for weak disorder in the small wavelength regime. The two terms in \Eq{eLOC} correspond to on-diagonal and off-diagonal disorder respectively. We discuss the two terms separately below.

The hermitian real matrix $\bm{H}$ is formally identical to that of the Anderson-Debye model, see Appendix~C.
The {\em off-diagonal disorder}, aka resistor network disorder, is the same type of disorder that appears e.g. in the Debye model (balls connected by springs). The strength of this disorder is defined as follows:
\beq \label{eSigOff}
\sigma_{\perp}^2 \ \equiv \ \bar{w}^4 \text{Var}\left[ \frac{1}{w_n} \right] 
\ \approx \ \frac{1}{4} w_o^2 e^{-f} \text{Var}[f_n]
\eeq 
This definition assumes that the bonds are uncorrelated. 
The approximation is based on first order treatment of the disorder in \Eq{ewn}. Looking at \Eq{eLOC} we see that its effect is significant for the short-wavelength modes, and can be neglected in the vicinity of $\lambda=0$. It is the same as in the Debye model where it is argued that the long-wavelength modes tend to be extended.  

The {\em on-diagonal disorder} is more subtle. In the standard Anderson model all bonds are identical and it is common to consider white on-site disorder. Here it is not the case, hence the definition for the strength of the disorder becomes more subtle:
\beq \label{eU}
\sigma_{\parallel}^2 \ \equiv \ \text{Var}_{\lambda} \left[ u_n-w_n- w_{n{+}1}  \right] 
\eeq   
If $\bm{H}$ were a stochastic kernel that preserve probability, then we would get from this expression ${\sigma_{\parallel}=0}$, and would be left with Debye-type localization only. We have extra diagonal disorder analogous to pinning of the balls to the ground in the Debye model. This extra disorder leads to Anderson-type localization at the vicinity of ${\lambda=0}$.  
The second issue to notice is the subscript in $\text{Var}_{\lambda}$. This subscript reminds us that the diagonal elements are not independent random variables. Consequently we do not have ``white disorder" and the variance has to be calculated at the ``energy" of interest. Namely, the Born approximation \Eq{eLOC} is based on evaluation of matrix elements $\BraKet{-k}{U(x)}{k}$ for backscattering. Here we use for clarity continuous space notations ($x$ instead of $n$). Averaging the squared matrix elements over realizations of the potential $U(x)$ one deduces that     
\beq
\text{Var}_{\lambda}[U(x)] \ = \ \sum_r e^{i k r} C(r) 
\eeq
where $C(r)$ is the correlation function of the disorder.
Here we are back with discrete notions, accordingly 
the distance $r$ between sites is an integer number.  
For ``white" disorder $C(r) = \text{Var}(U_n) \delta_{r,0}$.
But the potential $U_n$ in the square brackets of \Eq{eU} is correlated.    
For presentation purpose we assume also ${f \ll 1}$, 
which corresponds to the continuum limit,  
while the more general case is addressed in Appendix~D. 
The disorder in leading order comes out  
\beq 
U_n = - \frac{w_o}{2}(f_n-f_{n-1})-\lambda g_n + \text{const} 
\eeq
Consequently we obtain 
\beq 
\sigma_{\parallel}^2 \ \approx \ 
\frac{w_o}{4}\lambda \text{Var}[f_n]+\lambda^2 \text{Var}[g_n] 
\eeq
Substitution into \Eq{eLOC} we see, as anticipated, that off-chain disorder provides a contribution proportional to $\lambda$ that always vanishes in the vicinity of $\lambda=0$, while on-chain disorder does not vanish.

The bottom line is very simple, and we summarize it in simple words: the inverse localization length is determined by the effective diagonal disorder. The strength of this disorder is proportional to $\lambda^2$ for off-chain disorder, and therefore we always get a complex bubble at the vicinity of the origin, indicating under-damped relaxation. For on-chain disorder the  inverse localization length approaches a finite value at the limit ${\lambda \rightarrow 0}$. Therefore the complexity depends on the slope  $\kappa'(0)$ at the origin. This slope becomes negative for large enough $f$, hence we get a delocalization transition. The details in the latter case are the same as in the ``standard model", see \cite{psl} where also the complexity saturation is explained.

\section{Summary}

Stochastic networks are of general interest in many fields of Physics, 
as well as in Chemistry and Engineering. Key questions in the study of such 
networks are how they respond to bias, and what are their relaxation modes. 
In the traditional studies of tight binding models the main observations 
have to do with the sliding and the delocalization transitions. 
Once we allow more complex quasi-one-dimensional configurations, some new issues emerge. 

The delocalization of the relaxation modes for the ANM configuration of \Fig{f1}c has been already studied in~\cite{ndm}. In the present work we have focused on the NDM configuration of \Fig{f1}b, which is the simplest version of a comb-type model \cite{havlin,balakrishnan,zia,benarous}.

In the preliminary pedagogical sections we used the conventional NESS perspective in order to derive the dependence of the steady state current on the bias (see e.g. \Fig{f2} and \Fig{f3} and \Fig{f4}). Then we clarified  that results for~$v$ and~$D$ can be regraded as spectral properties that characterize the relaxation spectrum. From this point on our interest has been focused on the study of this spectrum, and specifically in the delocalization transition of the eigen-modes.  

The study was partially  motivated by the following question: we know that in one dimension we always have localization; does it mean that in a closed ring we always have a delocalization transition? It was essential in this context to distinguish between on-chain disorder and off-chain disorder. In particular we found that off chain disorder leads to localization that is not strong enough to induce over-damped relaxation, hence delocalization transition is absent. 
It is implied that for off-chain disorder also sliding transition does not take-place. In fact the absence of sliding transition is much easier for understanding using a NESS-perspective because for off-chain disorder activation barriers are not formed.                  

On the formal side we have explained how the analysis of the relaxation spectrum  
can be carried out using a reduced tight binding model. 
Thus the relaxation spectrum (eigenvalues of $\bm{\mathcal{W}}$) can be related  
to the real spectrum of a real Hermitian matrix $\bm{H}$ that describes 
an effective Anderson-Debye model. In order to figure out the delocalization transition,  
the $\lambda$ dependence of the inverse localization length is required. This dependence is very different for on-chain and off-chain disorder.

\ \\
\sect{Acknowledgment}
This research was supported by the Israel Science Foundation (Grant No.283/18).

\clearpage
\onecolumngrid
\appendix

\section{The Bloch matrix}


For a one dimensional chain with dangling bonds, in the absence of disorder, 
the matrix $\bm{\mathcal{W}}$ can be written using momentum and spin operators.
For symmetric transitions  
\beq
\bm{\mathcal{W}} \ = \ c_o[\bm{\sigma}_x-1] + \sum_{\pm}\bm{\sigma}_{\uparrow}w_o[e^{\pm i\mathbf{P}}-1]
\eeq
where the momentum operator is defined such that ${e^{\mp\bm{P}}\ket{n} = \ket{n{\pm}1}}$, and ${\bm{\sigma}_{\uparrow} = \frac{1}{2}[\bm{1}+\bm{\sigma}_{z}]}$ is a projector on the chain sites, while the Pauli operator $\bm{\sigma}_x$ induce transitions between $\uparrow$ and $\downarrow$ sites.  
The momentum with eigenvalue $k$ is a constant of motion, 
and therefore the matrix decomposed into blocks: 
\beq 
\bm{\mathcal{W}}^{(k)} = 
\left(\amatrix{
-c^{+} + (e^{-ik}{-}1)w^{+} + (e^{ik}{-}1)w^{-}   & c^{-} \cr  c^{+} & -c^{-}
}\right) \ \ \ \ \ 
\eeq
In the latter expression we have assumed that transitions do not have 
the same rates in the forward and in the backward directions, 
and therefore replaced $c_o$ by $c^{\pm}$,  
leading to an asymmetric non-hermitian matrix that describes \Fig{f1}b.   
For the quasi one dimensional network of \Fig{f1}c we add the transitions 
along the $\downarrow$ sites and get
\beq 
\bm{\mathcal{W}}^{(k)} = 
\left(\amatrix{
 -c^{+}  + (e^{-ik}{-}1)w_{\uparrow}^{+} + (e^{ik}{-}1)w_{\uparrow}^{-} & c^{-} \cr  
c^{+} &  - c^{-} + (e^{-ik}{-}1)w_{\downarrow}^{+} + (e^{ik}{-}1)w_{\downarrow}^{-} 
}\right) \ \ \ \ \ 
\eeq
The diagonaliation of $\bm{\mathcal{W}}^{(k)}$ provides the $\lambda_{\nu}(k)$ spectrum, from which analytical expressions for the drift velocity and the diffusion coefficient are derived.

\section{Band boundaries}

The expressions for the band boundaries, assuming $w_o=c_o=1$ are: 
\beq
\lambda_{0}(\pi) &=& \frac{1}{2}e^{-(1+\alpha)f}\left[ e^{f}+2e^{\alpha f}+3e^{(1+\alpha)f} 
+\sqrt{ e^{2f}+4e^{2\alpha f}+9e^{2(1+\alpha)f}-2e^{(2+\alpha)f}-4e^{(1+\alpha)f}+12e^{(1+2\alpha)f} } \right]
\\
\lambda_{1}(0) &=& 1 + e^{-\alpha f} 
\\
\lambda_{1}(\pi) &=& \frac{1}{2}\left[ 3+2e^{-f}+e^{-\alpha f} 
-e^{(1+\alpha)f}\sqrt{e^{2f}+4e^{2\alpha f}+9e^{2(1+\alpha)f}-2e^{(2+\alpha)f}-4e^{(1+\alpha)f}+12e^{(1+2\alpha)f} } \right]
\eeq
The full expression for $\lambda_0(k)$ is not too illuminating, and therefore is not displayed. Its second derivative at ${k=0}$ gives the diffusion coefficient $\mathcal{D}$ of \Eq{eD} which is plotted in \Fig{f2}b. 
For completeness we write also what is result for the diffusion coefficient for the active network:
\beq
\mathcal{D} \ = \
\left[ \frac{ (1 + e^{\alpha f-f}) + e^{-\phi} (e^{-f} + e^{\alpha f}) } {1+e^{\alpha f}} \right] \frac{w_o}{2} 
+ \left[ \frac{ e^{2\alpha f}  (1+e^{-f})^2 (1-e^{-\phi})^2 } {(1+e^{\alpha f})^3}  \right] \frac{w_o^2}{c_o}
\eeq

\begin{figure}[t!] 
\centering
\includegraphics[height=4cm]{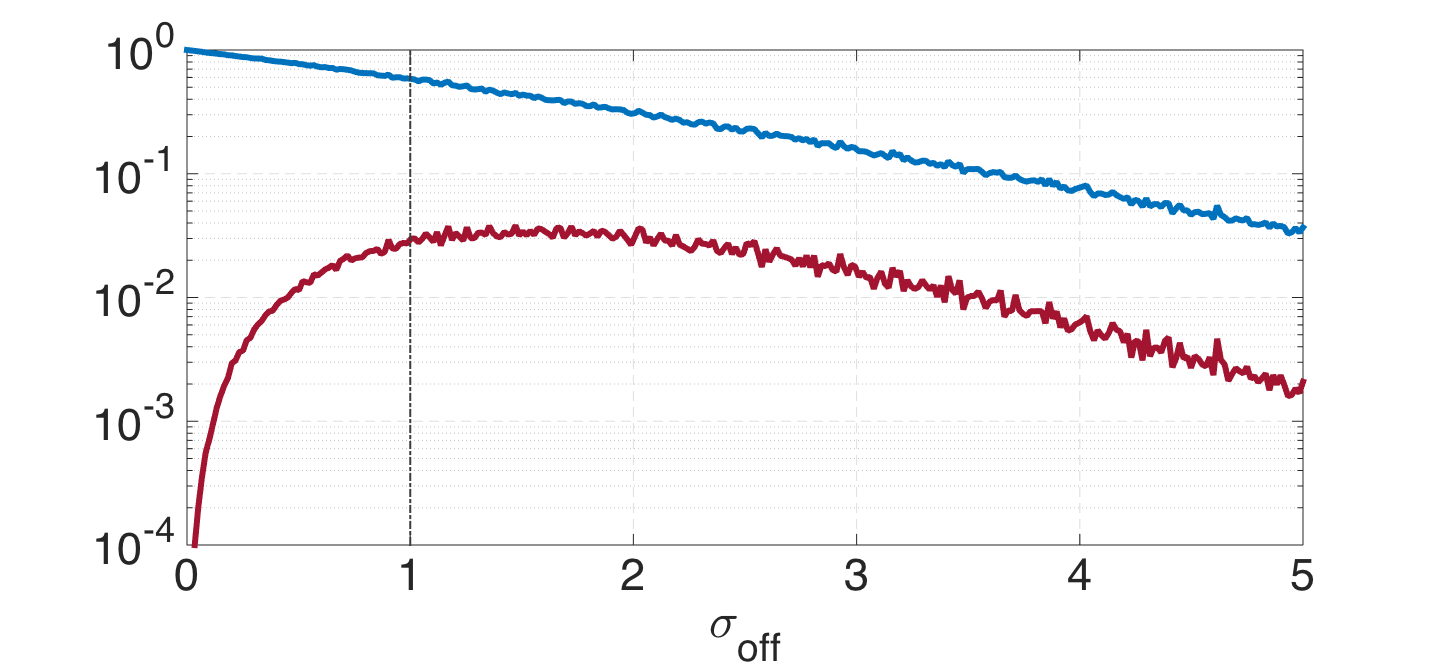}
\includegraphics[height=4.5cm]{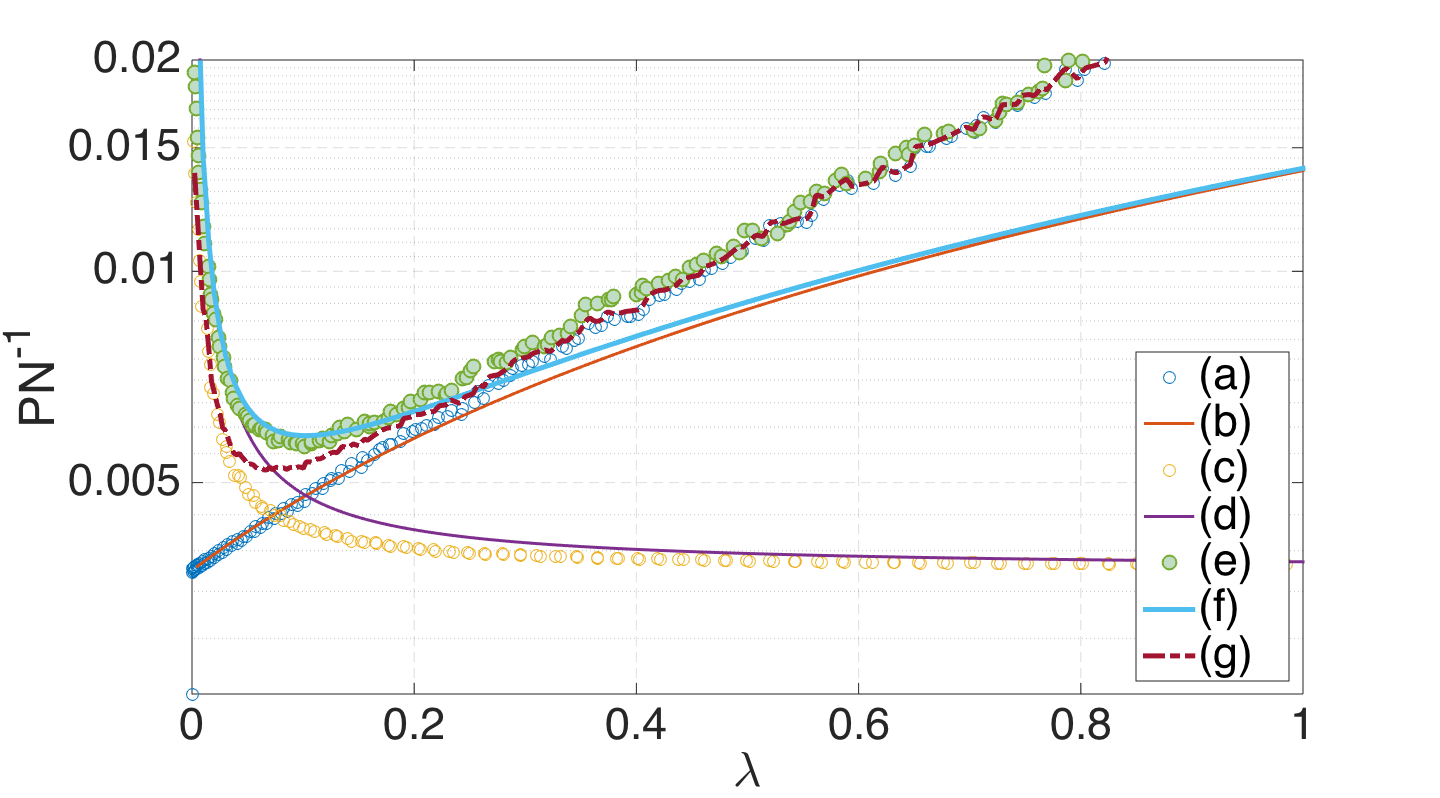} 
\caption{{\bf The perturbative estimate of the PN.} 
Left panel: The dependence of $\bar{w}$ (upper, blue) and $\sigma_{\perp}$ (lower,red) on ${ \sigma_{\text{off}} }$. We use harmonic average for the former, and \Eq{eSigOff} for the latter.   
Right panel: The PN is numerically determined for a ring of $L{=}400$ sites, based on 500 realizations of the disorder. Plots (a) and (c) are for off-diagonal disorder ${\sigma_{\text{off}} = 1}$ 
and for on-diagonal disorder ${\sigma_{\text{on}} = 0.05}$, respectively. Plot (e) is what we get if we have both. Plot (g) corresponds to $\kappa_{\text{off}}+\kappa_{\text{on}}$, based on (a) and (c). Curves (b,d,f) are the analytical estimates based on \Eq{eLOC} with no fitting parameters.   
}  
\label{fPNvsFGR}
\end{figure}

\section{The Anderson-Debye model}

What we call Anderson-Debye model refers here to a system whose dynamics is dictated by a real symmetric matrix $\bm{H}$ that describes a one dimensional tight binding chain with both diagonal disorder and off-diagonal disorder, namely,
\beq
\bm{H} \ = \ \sum_n \ket{n}u_n\bra{n}  + \sum_{n} \Big[\ket{n} w_{n} \bra{n{-}1} + \ket{n{-}1} w_{n} \bra{n}\Big]
\eeq 
The Anderson model arises in the context of electronic conduction: a particle hops with frequency $w_n$ between sites that do not have the same potential $u_n$. The Debye model refers to balls that are connected by springs that have spring-constants $w_n$, and that are possiblly grounded to the floor with extra springs (aka pinning). If the disorder is weak we define 
\beq
\sigma_{\perp}^2 \ &\equiv& \ \text{Var}\left[ w_n \right] \\ 
\sigma_{\parallel}^2 \ &\equiv& \ \text{Var}\left[ u_n-w_n-w_{n{+}1} \right] 
\eeq
In the absence of pinning ($\sigma_{\parallel}=0$) Debye has correctly conjectured that the low lying excitations (${\lambda \rightarrow 0}$) are extended free waves. Otherwise the eigenstates are localized, as argues by Anderson. Leading order perturbation theory (the Fermi-golden-rule, aka in this context the Born approximation) leads to \Eq{eLOC} for the inverse localization length.

A slightly improved version of \Eq{eLOC}, see \cite{rna}, 
involves the harmonic average $\bar{w}$ for the mean coupling, and \Eq{eSigOff} for its dispersion. The rational of looking on the statistic of $1/w_n$ is based on the formal analogy with a resistor-network (bonds added in series), where rigorously $1/\bar{w}$ is the resistivity of the bond-disordered chain.  

The limitations of the perturbative expression \Eq{eLOC} have been pointed out in the main text. For demonstration purpose we test its usefulness in \Fig{fPNvsFGR}. We consider a chain of length $L$ with 
\beq
w_n \ &=& \ e^{-\text{random}[0,\sigma_{\text{off}}]} \\
u_n \ &=& \ w_n+w_{n{+}1} + \text{random}[-\sigma_{\text{on}},\sigma_{\text{on}}] 
\eeq 
For a non-disordered sample we have $\text{PN}_0 = L$, while for a disordered ring we expect ${\text{PN}^{-1} = \text{PN}_0^{-1} + \kappa}$, see \cite{kottos}. 
We test both the additivity of $\kappa$ which is implied by \Eq{eLOC}, and also the formula as it is, with no fitting parameters. We conclude that it works reasonably well in the $\lambda$ range of interest.

\section{The on-diagonal disorder}

We write the bias for a chain bonds $f_n = f+\tilde{f}_n$, 
and for a dangling bonds $g_n = \alpha f + \tilde{g}_n$. 
Accordingly we have for weak disorder, after dropping a constant,  
\beq
U_n \ \approx \ -\frac{w_o}{2}e^{-f/2} 
\left[ (2e^{-f/2}-1)\tilde{f}_n -  \tilde{f}_{n{-}1} \right] 
- \lambda  e^{\alpha f} \tilde{g}_n
\ \equiv \ A \tilde{f}_n + B \tilde{f}_{n{-}1} + C \tilde{g}_n 
\eeq
Consequently we get for the effective strength of the disorder 
\beq
\sigma_{\parallel}^2 \ \approx \ 
[A^2+B^2+ 2AB\cos(k)]\text{Var}[f_n] + C^2 \text{Var}[g_n] 
\eeq
In the main text we have highlighted the continuum limit (${f \ll 1}$) 
for which ${A\approx-B}$, and ${\cos(k)\approx 1 - (1/2)k^2}$, 
and therefore the first term is proportional to $\lambda$. 
In general we might have a term that does not vanish 
in the limit ${\lambda \rightarrow 0}$. Then one has to use a formula 
that goes beyond the diverging approximation of \Eq{eLOC}.


\clearpage
\end{document}